\documentclass[a4paper,11pt]{article}
\usepackage{jheppub}

\usepackage{graphicx}   
\usepackage{dcolumn}    
\usepackage{bm}         
\usepackage{hyperref}   
\hypersetup{
  colorlinks=true,
  linkcolor=blue,
  citecolor=blue,
  urlcolor=blue,
}
\usepackage{diagbox}
\newcommand{\be}{\begin{equation}}
\newcommand{\ee}{\end{equation}}
\newcommand{\ba}{\begin{aligned}}
\newcommand{\ea}{\end{aligned}}
\renewcommand{\d}{\partial }
\DeclareMathOperator{\Tr}{Tr}
\DeclareMathOperator{\tr}{tr}
\def\Im{\mathop{\mathrm{Im}}\nolimits}

\def\Re{\mathop{\mathrm{Re}}\nolimits}
\def\mb{\mathbb}

\def\mc{\mathcal}

\def\bp{\begin{pmatrix}}
\def\ep{\end{pmatrix}}

\usepackage{comment} 

\title{Constraining F-theory Model Building with QCD Axions}

\author{Keren Chen$^1$}

\author{Qinjian Lou$^1$}

\author{Yi-Nan Wang$^{1,2}$}

\affiliation{%
 $^1$School of Physics, Peking University 
}%

\affiliation{$^2$Center for High Energy Physics, Peking University}

\date{\today}
\emailAdd{keren.chen246@gmail.com,qinjian.lou@pku.edu.cn,ynwang@pku.edu.cn}
\abstract{
In this paper, we investigate axion physics in 4D F-theory MSSM models over base threefolds with low $h^{1,1}$. We derive the axion coupling to the QCD gauge fields and the axion potential from a top-down perspective, in both the IIB superstring and the dual M-theory picture. For the explicit geometric models, we employ the ``quadrillion'' landscape of 4D F-theory models with the exact Standard Model chiral spectrum, and study simple base threefolds with low $h^{1,1}$, namely $\mathbb{P}^3$, $\mathbb{P}^1\times\mathbb{P}^2$, the generalized Hirzebruch threefold $\tilde{\mathbb{F}}_3$ and $\mathbb{P}^1\times\mathbb{P}^1\times\mathbb{P}^1$. We derive exclusion constraints on the K\"ahler moduli space of the base threefold from the CP violation angle, the Standard Model gauge coupling constants and the stretched K\"{a}hler cone condition. We find stringent constraints on the set of base divisors that should be rigid or rigidified by the inclusion of flux. For the allowed regions of the parameter space, we estimate the typical mass of detectable QCD axions to be around $10^{-9}$ eV, and the axion decay constant to be around $f_a\sim 10^{15}$ GeV. We emphasize that these estimates are specific to the low-$h^{1,1}$ bases studied here and are not expected to be generic in the F-theory landscape.}

\begin{document}
\maketitle

\section{Introduction}
The axion is a new particle commonly introduced in physics beyond the Standard Model in order to solve the strong CP problem in QCD~\cite{Peccei:1977hh,Peccei:1977ur,Nelson:1983zb,Peccei:2006as,Kim:2008hd,Reece:2023czb,Weinberg:1977ma,Wilczek:1977pj}. Axions (axion-like-particles) can also produce a variety of phenomena with significant implications for cosmology and particle physics~\cite{Preskill:1982cy,Kim:1986ax,Banks:2002sd,Kawasaki:2013ae,Marsh:2015xka,Irastorza:2018dyq,DiLuzio:2020wdo,Choi:2020rgn,Kim:1979if,Shifman:1979if,Dine:1981rt,Zhitnitsky:1980tq,Abbott:1982af,Dine:1982ah,Vysotsky:1978dc,Berezhiani:1989fp,Berezhiani:1992rk,Khlopov:1998uj,Khlopov:1999tm,Khlopov:2004sc}, motivating extensive searches for axions. Extensive experimental efforts have ruled out portions of the axion parameter range~\cite{AxionLimits}. In string theory, axions naturally appear in generic string compactification models over the compact space $X$~\cite{Choi:1985je,Banks:1996ea,Banks:1996ss,Conlon:2006tq,Svrcek:2006yi,Choi:2006qj,Grimm:2007hs,McAllister:2008hb,Arvanitaki:2009fg,Arvanitaki:2010sy,Acharya:2010zx,Panda:2010uq,Cicoli:2012sz,Ringwald:2012cu,Kamionkowski:2014zda,Bachlechner:2014gfa,Long:2016jvd,Visinelli:2018utg,Demirtas:2018akl,Demirtas:2021gsq,Cicoli:2022fzy,Gendler:2023kjt,Agrawal:2024ejr,Loladze:2025uvf,Cheng:2025ggf,Petrossian-Byrne:2025mto,Benabou:2025kgx,Leedom:2025mlr,Reig:2025dqb,Agrawal:2025rbr,Jain:2025vfh}. For example, there are multiple NS-NS and R-R sector $p$-form gauge fields in type II superstring theory. Integrating such form fields over compact cycles on $X$ leads to a large number of periodic scalars in 4d. Such scalars can be regarded as axions if they couple to the QCD $SU(3)$ gauge group, and axion-like-particles if they do not.

Nonetheless, the vast literature on the string axiverse is mostly set in weakly coupled superstring theory. It is natural to extend the discussion of axion physics to strongly coupled frameworks of superstring theory, i.e. F-theory as a geometric description of strongly coupled IIB superstring theory and a natural UV complete quantum gravity framework to realize Standard Model~\cite{Vafa:1996xn,Morrison:1996na,Morrison:1996pp,Beasley:2008dc,Beasley:2008kw,Donagi:2008ca,Grimm:2010ks,Taylor:2015xtz,Weigand:2018rez}. In the F-theory setups, axions were constructed and discussed using complex structure moduli~\cite{Grimm:2014vva,Grimm:2020ouv,Grimm:2025cpq} as well as the more conventional, imaginary part of the complex K\"{a}hler moduli~\cite{Grimm:2015zea}. Recently, the axiverse in F-theory was studied in detail, on elliptic Calabi-Yau fourfolds with large $h^{1,1}$ and $10^4\sim 10^5$ axion-like-particles~\cite{Fallon:2025lvn}, including the geometry with the largest $h^{1,1}$ constructed in \cite{Wang:2020gmi}.

In this paper, we analyze the parameter space and physics of axions in MSSM like models in string theory, with $SU(3)\times SU(2)\times U(1)_Y$ gauge groups and proper matter representations as in the Standard Model~\cite{Lin:2014qga,Cvetic:2015txa,Cvetic:2019gnh}. In particular, we employ the Weierstrass model used in the ``quadrillion'' set of F-theory MSSM models~\cite{Cvetic:2019gnh}, and analyze the axion physics which naturally appears in such models, over simple examples of base threefolds $B_3$ with small $h^{1,1}(B_3)$.

Axions (axion-like-particles) $a_i$ in F-theory setup naturally arise from the integration of the IIB $C_4$ gauge field over 4-cycles $D_i$ on the base threefold $B_3$. Thus, their kinetic terms descend from the kinetic terms of $C_4$ in 10D IIB supergravity. From the perspective of the 7-brane worldvolume action, the topological term $C_4\wedge\tr(F\wedge F)$ naturally produces the axion coupling term $a\tr(F\wedge F)$ in 4D. We also confirm the presence of this term from an argument of M/F-duality, in the dual M-theory picture.

For the axion potential, in this work, we first consider the 4D $\mc{N}=1$ superpotential (\ref{W-total}) with the tree-level Gukov-Vafa-Witten term $W_0$ and quantum corrections from Euclidean D3-branes (E3) wrapping 4-cycles $D$ on the base $B_3$. Importantly, the presence of the subleading quantum corrections, which are also important in the stabilization of K\"{a}hler moduli, requires $D$ to be rigid, or rigidified by flux~\cite{Bianchi:2011qh,Bianchi:2012pn} (for example the divisors on $B_3=\mb{P}^3$). The resulting scalar potential represents the top-down contribution to the axion potential, denoted as $\mc{V}_{UV}$. We argue that the RG flow effect of $\mc{V}_{UV}$ to the IR is insignificant for our purposes, and it can be added up with the axion potential from IR QCD quantum effects, denoted as $\mc{V}_{IR}$. Solving for the minimum of the scalar potential $\mc{V}=\mc{V}_{UV}+\mc{V}_{IR}$ generally leads to a non-vanishing $\theta_{\text{QCD}}$ proportional to the vev of the axion, from the undetermined, $\mc{O}(1)$ complex phase of the Pfaffian $A$ in the non-perturbative superpotential. 

On the other hand, measurements of the neutron electric dipole moment constrain the CP violation angle to $|\theta_{\text{QCD}}|\lesssim10^{-10}$~\cite{Abel:2020pzs}. The smallness of $\theta_{\text{QCD}}$ sets a restriction on the axion potential contributed by the superpotential, which is determined by the moduli of the geometric setup. Hence the experimental constraints on $\theta_{\text{QCD}}$ can help us exclude some parameter regions of compact space. Along with the lower bounds on the value of the Standard Model gauge coupling constants interpolated to the high energy scale, and the 1-stretched K\"ahler cone condition to suppress stringy corrections to the supergravity approximation~\cite{Fallon:2025lvn}, many models with specific rigid (or rigidified) divisors in \cite{Cvetic:2019gnh} are excluded.

Since the UV Standard Model gauge coupling constants depend on the SUSY breaking scale, we have two sets of bounds. The first one is more restrictive, corresponding to an arbitrary SUSY breaking scale above TeV, and the geometric models satisfying such bound are called ``\texttt{Y}''. The second one corresponds to a high SUSY breaking scale, and we label the geometric models in this region as ``\texttt{O}''. Otherwise the model is excluded and is labeled as ``\texttt{N}''. For example, for the simplest base $B_3=\mb{P}^3$, the physical allowed region labeled by ``\texttt{Y}'' requires that the rigidified divisor class is $[nH]$, where $H$ is the hyperplane class of $\mb{P}^3$ and $n>11$, while the less restrictive region labeled by ``\texttt{O}'' requires that $6<n\leq 11$. For $1\leq n\leq 6$, the model is fully excluded and is labeled as ``\texttt{N}''. Hence a highly constrained rigidification is mandatory for the F-theoretic MSSM in \cite{Cvetic:2019gnh}, over simple bases. 

Furthermore, we compute the predicted axion masses $m_a$ and inverse axion decay constants $1/f_a$ in several physically allowed examples, and they typically lie around $m_a\sim 10^{-9}$eV and $1/f_a\sim 10^{-15}$GeV$^{-1}$, see Fig.~\ref{m-f2}.

The structure of this paper is as follows: we review the strong CP problem and QCD axion effective action in section \ref{sec:QCD}. In section~\ref{sec:EFF} we present the derivation of axion effective action at the IIB superstring/F-theory level. We extract the axion potential from the scalar potential (\ref{uvpo}) from 4D $\mc{N}=1$ supergravity, with contributions from only the rigid and rigidified divisors. This term will break the CP symmetry in QCD. We provide constraints on the Standard Model gauge coupling constants at the energy scale of string compactification in section~\ref{sec:RG}. In section~\ref{sec:more} we briefly discuss the effective action of the cases with multiple axions.

In section \ref{section3}, we employ the $SU(3)\times SU(2)\times U(1)_Y$ MSSM framework presented in \cite{Cvetic:2019gnh}, with $U(1)_Y$ Mordell-Weil group. We further obtain the coupling between the axion and gauge groups from the M-theory dual picture, matching the IIB result. 

With this model, in section \ref{sec:results} we compute the dependence of $\theta_{\text{QCD}}$ and the gauge  coupling constants $\alpha_1$, $\alpha_2$, $\alpha_3$ on parameters in the K\"{a}hler moduli space, for base threefolds $B_3=\mb{P}^3$, $\mb{P}^1\times\mb{P}^2$, the generalized Hirzebruch threefold $\tilde{\mb{F}}_3$ and $\mb{P}^1\times\mb{P}^1\times\mb{P}^1$. In this section we use the simplified choice of line bundles $\mc{S}_7=\mc{S}_9=-K_B$ in \cite{Cvetic:2019gnh}. We plot exclusion graphs for specific choices of the rigid (or rigidified) divisors, leading to examples satisfying the ``\texttt{Y}'', ``\texttt{O}'' or ``\texttt{N}'' conditions.

In section~\ref{sec:Tuning}, we perform a more systematic scan over the different choices of the line bundles $\mc{S}_7$, $\mc{S}_9$ and of the rigid (or rigidified) divisors, and summarize the statistics of the resulting classification; the full results are provided in the associated data file. 

We conclude in section~\ref{conclusion}. In Appendix~\ref{app:data-files} we explain the format of the associated data file and the exact criteria for the ``\texttt{Y}'', ``\texttt{O}'' or ``\texttt{N}'' conditions used in this paper. In Appendix~\ref{sec:CY4}, we summarize the systematics of computing the Hodge numbers of the smooth elliptic Calabi-Yau fourfolds $Y_4$ for the models in \cite{Cvetic:2019gnh}; the order of magnitude of $h^{3,1}(Y_4)$ is around $\mc{O}(10^2)$.

\section{F-theory axion and its effective action}
\label{section2}

\subsection{QCD axion}
\label{sec:QCD}

In QCD, it is allowed to add a topological $\theta$-term to the most general renormalizable Lagrangian:
\be
L_{\theta}=\frac{\theta}{32\pi^2}F^{\mu\nu}_a \tilde{F}_{a\mu\nu}\,,
\ee
where $\tilde{F}_{a\mu\nu}=\frac{1}{2}\epsilon_{\mu\nu\rho\sigma}F^{\rho\sigma}_a$. We adopt the convention in which the 4d Yang--Mills action is
\be
S_{\rm {YM}}=-\frac{1}{4g^2}\int \mathrm{d}^4x F_{a\mu\nu} F^{\mu\nu}_a\,.
\ee
The presence of the free parameter $\theta$ explicitly breaks CP symmetry in the Standard Model. However, the current null results from neutron electric dipole moment experiments~\cite{Abel:2020pzs} impose a stringent constraint on the effective QCD CP-violating angle\footnote{Here we adopt the conservative bound $|\theta| \lesssim 10^{-10}$ rather than the most stringent numerical limit available in the literature. The experiments directly constrain the neutron electric dipole moment $d_n$, while translating this into a bound on $\theta$ relies on non-perturbative QCD calculations relating $d_n$ to $\theta$. Since these calculations carry theoretical uncertainties, we use the order-of-magnitude estimate $|\theta| \lesssim 10^{-10}$, which is widely adopted in the literature.},
\be
|\theta| \lesssim 10^{-10}\,.
\ee
To explain the unnatural smallness of $\theta$, one possible solution is to promote $\theta$ to a dynamical periodic pseudoscalar field, namely the QCD axion $a$, with the Lagrangian
\be
L_a=\frac{1}{32\pi^2}\frac{a}{f_a}F^{\mu\nu}_a \tilde{F}_{a\mu\nu}-\frac{1}{2}\d_\mu a\d^\mu a\,.
\label{QCDaxion}
\ee
Here $f_a$ is the axion decay constant. Non-perturbative QCD effects generate an IR axion potential, which can be expressed in terms of the pion mass $m_\pi$, the pion decay constant $f_\pi$, and the up- and down-quark masses $m_u$ and $m_d$:
\be
\ba
V_{IR}=-m_{\pi}^2 f_{\pi}^2\sqrt{1-\frac{4 m_um_d}{(m_u+m_d)^2}\sin^2\left(\frac{a_{\text{QCD}}+\theta f_a}{2f_a}\right)}\,.
\ea
\label{pion}
\ee

Minimizing this IR potential dynamically leads to $\langle\theta + a/f_a\rangle=0$, thereby providing a solution to the strong CP problem. Moreover, the axion acquires a mass from this potential~\cite{Marsh:2015xka}:
\be
m_{a,QCD}\approx 6 \times 10^{-6}\text{eV}\times\left(\frac{10^{12}\text{GeV}}{f_a}\right)\,.
\label{qcdaxionmass}
\ee
Including the interactions with Standard Model gauge bosons, the effective Lagrangian becomes~\cite{Bauer:2020jbp}
\be
\ba
L_{eff}=&-\frac{1}{2}\d_\mu a\d^\mu a -\frac{m^2_{a,0}}{2}a^2-\frac{\d^\mu a}{f}\sum_F \bar{\psi}_F c_F\gamma_\mu \psi_F
+c_{GG}\frac{\alpha_3}{4\pi}\frac{a}{f}G^{a}_{\mu\nu}\tilde{G}^{\mu\nu,a}\\
&+c_{WW}\frac{\alpha_2}{4\pi}\frac{a}{f}W^{A}_{\mu\nu}\tilde{W}^{\mu\nu,A}
+c_{BB}\frac{\alpha_1}{4\pi}\frac{a}{f}B_{\mu\nu}\tilde{B}^{\mu\nu}\,.
\ea
\ee
We now discuss the renormalization group (RG) flow properties of the axion action and its couplings. $f$ is a constant, and the coefficients $c_{GG}$, $c_{BB}$, and $c_{WW}$ are invariant under RG flow ~\cite{Bauer:2020jbp,Chetyrkin:1998mw,Choi:2021kuy}. Therefore, the axion decay constant does not receive significant corrections when running from the string scale down to the GeV or MeV scale. This motivates us to compute $f_a$ ($f_a=\frac{f}{c_{GG}}$) directly at the string scale.

In an F-theory realization, one would like to include a UV contribution to the axion potential from a top-down perspective, in addition to the IR potential in Eq.~(\ref{pion}). We will introduce this UV potential in the next section.

\subsection{F-theory effective action}
\label{sec:EFF}

Similar to general string compactification scenarios, many axion-like particles naturally emerge in F-theory from $p$-form fields integrated over compact cycles. In particular, our objective is to realize the QCD axion coupled to Standard Model gauge fields and to derive the axion potential from the high-energy effective supergravity action (hereafter referred to as the UV action). With this potential, we can extract both the axion mass $m_{\text{axion}}$ and a non-zero CP-violating angle $\theta$.

We use the following string units for the equations in this paper:
\be
2\pi\sqrt{\alpha'}=l_s=1\,.
\ee
By matching the 10D Einstein-Hilbert action\ to its 4D counterpart upon integrating out the internal manifold, we have
\be
\ba
S_{4D}&=\frac{1}{16\pi G} \int \mathrm{d}^4 x \sqrt{-g} R\,, \\
S_{10D}&=\frac{2\pi}{l_s^8} \int \mathrm{d}^{10}x \sqrt{-g}R=2\pi V_{b}/l_s^8 \int \mathrm{d}^{4}x \sqrt{-g}R\,,\\
m_{p}&=\sqrt{\frac{\hbar c}{G}}=1.22089\times 10^{28}\mathrm{eV}\,, \\
l_s &=1.456\times 10^{-27} \sqrt{V_b}/l_s^3\mathrm{eV}^{-1}=1\,,
\ea
\ee
where $m_p$ the 4D Planck mass and $V_b$ the volume of compact internal space.

In Type IIB superstring theory setup, we realize a gauge group $G$ from a specific configuration of D$p$-branes. The action of gauge field arises from the DBI action and Chern-Simons (Wess-Zumino) action of a D$p$-brane in the string frame~\cite{Blumenhagen:2013fgp}. The Yang-Mills kinetic term can be extracted from the DBI action via expansion:
\be
S_{g}=-\mu_p \int \mathrm{d}^{p+1} x \sqrt{-g} g_s^{-1}\frac{1}{4} (2\pi \alpha')^2F_{\alpha\beta}F^{\alpha\beta}
\ee  
where $\mu_p \equiv T_p =  2\pi (4\pi^2\alpha')^{-\frac{p+1}{2}} = 2\pi l_s^{-(p+1)}$ for the D$p$-brane, and $F$ represents the field strength of gauge fields on the D$p$-brane. We should transform this action to the Einstein frame, to express the physical action of the Standard Model.
In the Einstein frame, the metric relation is $g_{\mu\nu,\text{s}} = g_s^{\frac{1}{2}} g_{\mu\nu, \text{E}}$, and the Yang-Mills action is transformed to
\be
S_{g}=-2\pi l_s^{-(p+1)}\int \mathrm{d}^{p+1} x \sqrt{-g}\frac{1}{16\pi^2} F_{\alpha\beta}F^{\alpha\beta}\,.
\label{gaugefieldaction}
\ee
The Chern-Simons term for a stack of $n$ D$p$-branes is given by  \cite{Blumenhagen:2013fgp}:
\be
\ba
S_{CS}=&\mu_{p}\int (n C_{p+1}+2\pi\alpha'C_{p-1}\wedge \tr{F}\\
&+\frac{1}{2}(2\pi\alpha')^2 C_{p-3}\wedge[\tr F\wedge F+\frac{n}{48}(\tr R_T\wedge R_T-\tr R_N \wedge R_N)] +...)\,.
\ea
\ee
Since the CS term is topological and independent of the metric $g_{\mu\nu}$, it retains the same form in both the string frame and the Einstein frame. Notably, there is a term in the CS action giving rise to the coupling between axions and instantons after the dimensional reduction to 4D:
\be
S_{\text{axion}}=\frac{\mu_7 (2\pi\alpha')^2}{2}\int C_{4}\wedge \tr (F\wedge F)\,.
\label{axion_ac}
\ee

In our setup, we consider F-theory, a non-perturbative formulation of Type IIB compactifications with general $[p,q]$ 7-branes back-reacting on the geometry \cite{Weigand:2018rez}. In F-theory, the axion-dilaton $\tau = C_0 + i g_s^{-1}$ is a varying function on the internal base manifold $B_3$, thus the string coupling $g_s$ is no longer a constant. Furthermore, there are specific monodromies around the 7-brane loci, and $\tau$, $g_s$ may become singular on the 7-branes. Consequently, the complete DBI action cannot apply to 7-branes in F-theory. However, since Eqs.~(\ref{gaugefieldaction}) and (\ref{axion_ac}) are independent of $g_s$ and the CS term is topological, we posit that Eqs.~(\ref{gaugefieldaction}) and (\ref{axion_ac}) remain valid in F-theory (see \cite{Heckman:2022muc,Tian:2024dgl}). 

In this framework, a stack of 7-branes fills the 4D Minkowski space and wrap a 4-cycle with volume $V_{\text{4-cycle}}$ in the compact space. The Standard Model gauge group and matter fields are realized as open string modes on 7-branes. From (\ref{gaugefieldaction}), we can obtain the gauge group action for $U(1)$ gauge fields
\be
S_{g}=-2\pi\int d^4 x \sqrt{-g} V_{\text{4-cycle}} \frac{1}{16\pi^2} F_{\mu\nu}F^{\mu\nu}\,.
\label{guagefield}
\ee
and non-abelian gauge fields with the normalization of Lie algebra generators $\{T^a\}$ as $\tr (T^a T^b)=\frac{1}{2}\delta^{ab}$:
\be
S_{g}=-2\pi\int d^4 x \sqrt{-g} V_{\text{4-cycle}} \frac{1}{8\pi^2} \tr(F_{\mu\nu}F^{\mu\nu})\,.
\label{guagefieldnab}
\ee
Thus we find the gauge coupling $g$ is related to the volume of 4-cycle as
\be
\ba
g^2&=\frac{2\pi}{V_{\text{4-cycle}}}\,, \\
\alpha &=\frac{g^2}{4\pi}=\frac{1}{2V_{\text{4-cycle}}}\,.
\ea
\label{gaugecouplingcst}
\ee
in the string unit. Now suppose that we define the axion $a$ from $C_4$ integrating over a 4-cycle $D$ on the base $B_3$:
\be
a=\int_{D}C_4\,.
\ee

In the string unit, the instanton coupling term with the gauge field descends from the action (\ref{axion_ac}): 
\be
\ba
\label{IIB-axion-coup}
S_{\text{axion}}&=\frac{1}{4\pi}\int_{\mb{R}^{1,3}} a \tr (F\wedge F)\,.
\ea
\ee
When there are multiple 4-cycles $D_i$s $(i=1,\dots,h^{1,1}(B_3))$ on the base $B_3$, we define the corresponding axions as
\be
\label{D-axion}
a_i=\int_{D_i}C_4\,.
\ee

For the full action of axion, we investigate the effective action of 4D F-theory~\cite{Grimm:2010ks}:
\be
\ba
S=&-\int \frac{1}{2} R_4 *1+K_{I\bar{J}}\mathcal{D}M^I\wedge*\mathcal{D}\bar{M}^{\bar{J}}+\frac{1}{2}\Re f_{\Lambda\Sigma}F^{\Lambda}\wedge *F^{\Sigma}+\frac{1}{2}\Im f_{\Lambda\Sigma}F^{\Lambda}\wedge F^{\Sigma}+\mc{V}*1\,.
\ea
\label{kin}
\ee
Here $K$ is the K\"ahler potential and $K_{I\bar{J}}=\partial_I\bar{\d}_{\bar{J}}K(M,\bar{M})$. $M$ denotes moduli fields which include the K\"ahler moduli. $\mathcal{D}$ denotes the covariant derivative, which includes the gauge connection whenever the corresponding chiral multiplet is charged under a gauge symmetry. In F-theory~\cite{Grimm:2010ks}
\be
K=-2\ln V_b-\ln\left[\int_{Y_4}\Omega\wedge \bar{\Omega}-\int_S \Tr(\phi'\wedge\bar{\phi}')\right]
\ee
where $Y_4$ is the resolved elliptic Calabi-Yau fourfold. $S$ is the 4-cycles on which the 7-brane wraps. $\phi'$ is a (2,0) form on the 4-cycle. As before, $V_b$ is the volume of the complex base threefold $B_3$.

The K\"ahler moduli $T_i$ of the base threefold $B_3$ ($i=1,\dots,h^{1,1}(B_3)$) in F-theory can be expressed as
\be
\ba
T_i&=\frac{1}{2}\int_{D_i}J_b\wedge J_b+d_{i ab}(z)\Tr(N^a\Re N^b)+i\int_{D_i}C_4=V_{D_i}+d_{i ab}(z)\Tr(N^a\Re N^b)+ia_i.
\ea
\ee
$V_{D_i}$ is the volume of the divisor $D_i$. The $N^a$ corresponds to $h^{2,1}(Y_4)-h^{2,1}(B_3)$ Wilson lines' degrees of freedom. We would not consider these degrees of freedom in this work, thus we have the IIB equation
\be
T_i=V_{D_i}+ia_i\,.
\ee
The imaginary part $a_i$ is exactly the axion corresponding to the divisor $D_i$, as defined in (\ref{D-axion}).

Hence the kinetic term of axions is given by
\be
S_{kin}=-\int \frac{\d^2 K}{\d T_i\d \bar{T}_j}\mathcal{D}T_i\wedge*\mathcal{D}\bar{T}_j\,.
\label{axionkinetic}
\ee

$\mc{V}$ in (\ref{kin}) is the scalar potential determined by the superpotential \cite{Grimm:2010ks}, as the sum of F-term and D-term. The D-term vanishes in general, and the F-term is expanded as
\be
\label{VF}
\mc{V}_F=e^K (K^{I\bar{J}}D_I W D_{\bar{J}} \bar{W}-3|W|^2)\,,
\ee
where $D_I=\frac{\d}{\d M_I}+\frac{\d K}{\d M_I}$ and $M_I$ are the moduli fields in chiral multiplets. We would assume that the complex structure moduli are already fixed at a higher energy scale, and only consider the dependence of physical quantities on the K\"ahler moduli. 

The superpotential $W$ is expressed as
\be
\label{W-total}
W=W_0+\sum_{D}A_{D}e^{-\frac{2\pi}{c_D}T_D}\,.
\ee
$W_0$ is the Gukov-Vafa-Witten flux superpotential~\cite{Gukov:1999ya}:
\be
W_0=\int_{Y_4} G_4\wedge \Omega_4\,.
\ee
$\Omega_4$ is the holomorphic (4,0) form on $Y_4$, and $G_4$ is the 4-form flux in the M-theory dual picture. The $W_0$ resembles an intersection number. We work in the regime of a generic F-theory compactification, with intermediate values of $G_4$ and $\Omega$. Hence $W_0$ can be estimated as an $\mathcal{O}(1)$ number in our analysis, see e.g. \cite{Denef:2005mm}.  

The second term in (\ref{W-total}) is the instanton correction from the gaugino condensation or E3-instanton. $T_D$ is the K\"ahler moduli of the divisor $D$ involved in the sum. 
For a divisor $D=\sum_i [n_D^i D_i]$, the corresponding K\"ahler moduli $T_D=\sum_in_D^i T_i$. $c_D$ is the dual Coxeter number of the gauge group on this divisor if it is wrapped by 7-branes. If no 7-brane wraps, $c_D=1$ and this term becomes $A_D e^{-2\pi T_D}$. The prefactor $A_D$ is the Pfaffian dependent on other moduli, i.e. the complex structure moduli and D3-brane moduli in the setup \cite{Witten:1996bn,Ganor:1996pe,Witten:1999eg,Harvey:1999as,Denef:2008wq,Bianchi:2011qh,Cvetic:2012ts,Alexandrov:2022mmy,Kim:2022uni}. Besides, the Pfaffian $A_D$ is non-vanishing only if the divisor is rigid or rigidified~\cite{Witten:1996bn,Bianchi:2011qh,Bianchi:2012pn,McAllister:2024lnt}. Recall that a divisor $D$ is rigid if 
\be
h_{+}^{\bullet}(D,\mathcal{O}_D)=(1,0,0), \quad h_{-}^{\bullet}(D,\mathcal{O}_D)=0
\ee
If $D$ is rigid and smooth, it supports two zero modes. Thus, Euclidean D3-branes wrapping on $D$ can contribute to the superpotential. Furthermore, some specific choice of fluxes can change the number of zero modes of a divisor so that this divisor can contribute a term to the superpotential. This divisor is rigidified by fluxes. On the other hand, a rigid divisor can also be de-rigidified. In general, the determination of which $G_4$ can rigidify which divisors on $B_3$ is a hard problem~\cite{Bianchi:2011qh,Bianchi:2012pn}, as horizontal and remainder fluxes would be relevant~\cite{Braun:2014xka}. We would not carry out this calculation in detail, and leave the explicit analysis of $G_4$ and rigidification in a future work.

We consider the case with no Wilson line degree of freedom and a set of rigid (or rigidified) divisors $\{D_\alpha=\sum_i n_\alpha^i D_i\}$ that contribute to the superpotential. Keeping only the terms involving axions, the scalar potential (\ref{VF}) with (\ref{W-total}) can be expanded as
\be
\ba
\mc{V}=&e^{K}\sum_{ ij} K^{i\bar{j}}[\sum_{\alpha\beta}\frac{4\pi^2}{c_\alpha c_\beta}e^{-2\pi(n_\alpha^k\tau_k/c_\alpha+n_\beta^k\tau_k/c_\beta)}|A_\alpha A_\beta|\\
&n_\alpha^i n_\beta^j\cos (2\pi n_\alpha^k a_k/c_\alpha-2\pi n_\beta^k a_k/c_\beta-\phi_\alpha+\phi_\beta)\\
&+\sum_\alpha \frac{2\pi v_j n_\alpha^i|A_\alpha|}{c_\alpha V_b}e^{-2\pi n_\alpha^k\tau_k/c_\alpha}(|W_0|\cos(2\pi n_\alpha^k a_k/c_\alpha+\phi_0-\phi_\alpha)\\
&+\sum_\beta |A_\beta|e^{-2\pi n_\beta^k\tau_k/c_\beta}\cos(2\pi n_\alpha^k a_k/c_\alpha-2\pi n_\beta^k a_k/c_\beta+\phi_\beta-\phi_\alpha))].
\ea
\label{potential}
\ee 
Here $\phi_\alpha$ is the complex phase of $A_\alpha$, $\phi_0$ is the complex phase of $W_0$, and $K^{i\bar{j}}$ is the inverse of the K\"{a}hler metric $K_{i\bar{j}}=\partial_i\partial_{\bar{j}}K(M,\bar{M})$. We typically assume that the volume of the 4-cycles is sufficiently large; otherwise, the effective action may receive strong quantum gravity corrections. Thus, 
\be
e^{-2\pi n_\alpha^k\tau_k/c_\alpha}\ll1\,.
\label{expsmall}
\ee
To control the $\alpha'$ corrections, we require the moduli to reside within the stretched K\"ahler cone, similar to \cite{Fallon:2025lvn}. This implies that the volumes of all effective 2-cycles are not only positive, but they must also exceed a certain lower bound $r$ in string units. In this work, we set $r=1$, i.e. using the 1-stretched K\"ahler cone condition in \cite{Fallon:2025lvn}.

As for the Pfaffian $A_D$, which is determined by the other moduli, its explicit calculation is generally difficult. However, its value can be obtained in certain special cases \cite{Kim:2022uni}. For $\mathcal{N}=1$ super-Yang-Mills theory, the modulus of the gaugino condensate superpotential in $SU(N_{D7})$ super-Yang-Mills with ultraviolet cutoff $M_{\text{UV}}$ is given by \cite{Baumann:2006th}:
\be
|W_{np}|=16\pi^2M_{UV}^3\text{exp}\left(-\frac{T_3 V_D}{N_{D7}}\right)\,.
\ee
Here $T_3=2\pi$ in string units in our Einstein frame, and $N_{D7}$ is equal to the dual Coxeter number. The effective action is valid below the KK scale, which is related to the inverse length scale of the compact space $\sim V_b^{-1/6}$, and we take this scale as $M_{UV}$. If the volume of the compact space is $\sim 10^{3-4}$ in string units, we can estimate $16\pi^2M_{\text{UV}}^3$ as an $\mathcal{O}(1)$ factor. Thus, we crudely approximate $A_D$ as an $\mathcal{O}(1)$ factor. In the later analysis, we will show that our physical conclusions are insensitive to the value of the parameter $A_D$. With the relation (\ref{expsmall}), we have
\be
\ba
& 2\frac{\pi v_j n_\alpha^i|A_\alpha|}{c_\alpha V_b}e^{-2\pi n_\alpha^k\tau_k/c_\alpha}|W_0|\gg \\
 &\frac{4\pi^2}{c_\alpha c_\beta}e^{-2\pi(n_\alpha^k\tau_k/c_\alpha+n_\beta^k\tau_k/c_\beta)}|A_\alpha A_\beta|n_\alpha^i n_\beta^j\,,\\
 &2\frac{\pi v_j n_\alpha^i|A_\alpha|}{c_\alpha V_b}e^{-2\pi n_\alpha^k\tau_k/c_\alpha}|A_\beta|e^{-2\pi n_\beta^k\tau_k/c_\beta}\,.
\ea
\ee

Retaining only the leading-order term yields the axion's UV potential:
\be
\ba
\mc{V}_{\text{UV}}=&e^{K}\sum_{ij} K^{i\bar{j}}\sum_\alpha 2\frac{\pi v_j n_\alpha^i|A_\alpha W_0|}{c_\alpha V_b}e^{-2\pi n_\alpha^k\tau_k/c_\alpha}\cos(2\pi n_\alpha^k a_k/c_\alpha+\phi_0-\phi_\alpha)\,.
\ea
\label{uvpo}
\ee
To obtain the full potential of the axion in the IR, we must add the non-perturbative QCD quantum effects (\ref{pion}). Note that $\mc{V}_{\rm UV}$ and the QCD contribution originate from different effects: the former arises from Euclidean D3-branes (or gaugino condensation) on the rigid or rigidified divisors $D_\alpha$ and is generated at the compactification scale, while the latter arises from the non-perturbative dynamics of QCD in the IR. The two contributions are therefore added. Assuming the $SU(3)$ gauge group resides on a divisor represented by $D_G=\sum_{i} n_i D_i$, the coupling between axions and QCD instantons is expressed as a sum
\be
\label{mult-axion-coupling}
S_{\text{axion}}=\frac{1}{4\pi}\int_{\mb{R}^{1,3}} \sum_i(n_i a_i) \tr (F\wedge F)\,.
\ee
Comparing with (\ref{QCDaxion}), we identify the expression for the normalized QCD axion $(a_{\text{QCD}}/f_a)$ with $(2\pi\sum_i n_i a_i)$, thus the CP violation angle in our string setup is
\be
 \theta_{\text{QCD}}=\langle 2\pi\sum_i n_i a_i\rangle\,.
\label{string_angle}
\ee
Consequently, we obtain the IR potential contributed by the pion mass term in string units:
\be
\mc{V}_{IR}\approx \frac{1}{2}(2.6\times 10^{-38}V_b)^2\sin^2 (\pi \sum_{i}n_i a_i).
\label{uvpion}
\ee
The analysis can be extended analogously to derive the axion couplings to the other gauge groups, such as the Standard Model $SU(2)_L$ or $U(1)_Y$ gauge group.

Notice that the instanton coupling term is determined by the 7-brane Chern-Simons term, and we do not need to introduce the $\theta$ term as in (\ref{QCDaxion}). The full axion potential as seen for IR is $\mc{V}_{\text{UV}}+\mc{V}_{\text{IR}}$, from which we can calculate the mass $m$ of axions. Furthermore, the gauge coupling constant is determined by the volume of divisors wrapped by 7-branes hosting the Standard Model gauge groups, which also imposes constraints on the K\"{a}hler moduli space of $B_3$. We will discuss it in the following section.

\subsection{RG flow of the gauge coupling constants}
\label{sec:RG}

At one-loop level, the renormalization group equations for the gauge couplings $g_i$ can be written as
\begin{equation}
\mu \frac{d g_i}{d\mu} = \frac{b_i}{16\pi^2} g_i^3 \, ,
\end{equation}
where the coefficients $b_i$ depend on the particle content of the theory. For convenience, we will discuss the RG flow with the GUT normalization in this section, i.e. $g_1=\sqrt{5/3}g_Y$.

In the Standard Model (SM), using the $\overline{\text{MS}}$ scheme, the one-loop beta function coefficients are
\begin{equation}
(b_1,\, b_2,\, b_3)_{\text{SM}} = \left(\frac{41}{10},\, -\frac{19}{6},\, -7 \right).
\end{equation}

In the Minimal Supersymmetric Standard Model (MSSM), the coefficients become
\begin{equation}
(b_1,\, b_2,\, b_3)_{\text{MSSM}} = \left(\frac{33}{5},\, 1,\, -3 \right).
\end{equation}

The difference arises from the enlarged particle spectrum in the MSSM, where each Standard Model particle is accompanied by a superpartner. As a result, additional fermionic (gaugino, Higgsino) and scalar (sfermion) degrees of freedom modify the running of the gauge couplings, generally reducing the magnitude of asymptotic freedom and enabling gauge coupling unification at a high energy scale.

In our setup, the bulk volume is taken to be $V_b = \mathcal{O}(10\text{--}10^2)$ in string units. 
The Kaluza-Klein scale is therefore estimated as
\begin{equation}
M_{\mathrm{KK}} \sim \frac{M_s}{V_b^{1/6}} \, ,
\end{equation}
which typically gives $M_{\mathrm{KK}} \sim 10^{16}\,\text{GeV}$ for a string scale close to the Planck scale.
The 4D $\mc{N}=1$ supergravity action of section~\ref{sec:EFF}, in which the Kaluza-Klein modes have been integrated out, is only reliable sufficiently below the Kaluza-Klein threshold. Accordingly, we take the four-dimensional effective field theory description to be valid up to energies $\mu \lesssim 10^{14}\,\text{GeV}$, and we compare the geometric gauge couplings (\ref{gaugecouplingcst}) with the RG-evolved couplings at this scale rather than at $M_{\mathrm{KK}}$ itself.

The supersymmetry breaking scale is assumed to lie in the range
$M_{\text{SUSY}} \sim 10^{4}\text{--}10^{11}\,\text{GeV}$ in this paper. The lower end corresponds to a SUSY breaking scale just above the TeV scale, while the upper end is chosen such that $M_{\text{SUSY}}$ stays well below $\mu\sim 10^{14}\,\text{GeV}$: the derivation of the axion action from 4D $\mc{N}=1$ supergravity in section~\ref{sec:EFF} requires supersymmetry to be unbroken over a sufficiently wide range of energies below the Kaluza-Klein scale.

We apply the model presented in~\cite{Cvetic:2019gnh} to calculate the RG flow of the gauge coupling constants from the electroweak scale, where they are measured, up to $\mu\sim10^{14}\,\text{GeV}$. 

We evolve the gauge couplings from the electroweak scale to $\mu \sim 10^{14}\,\text{GeV}$ using the one-loop renormalization group equations
\begin{equation}
\frac{1}{\alpha_i(\mu)} =
\frac{1}{\alpha_i(M_Z)}
- \frac{b_i^{\mathrm{SM}}}{2\pi} \ln\frac{M_{\text{SUSY}}}{M_Z}
- \frac{b_i^{\mathrm{MSSM}}}{2\pi} \ln\frac{\mu}{M_{\text{SUSY}}} \, .
\end{equation}

Here $M_Z$ denotes the mass of the $Z$ boson, which sets the electroweak scale,
\begin{equation}
M_Z \simeq 91\,\text{GeV} \simeq 10^{11}\,\text{eV}.
\end{equation}
We use it as the reference scale where the gauge couplings are experimentally measured.

We take
\begin{equation}
\ba
&(b_1,b_2,b_3)_{\mathrm{SM}} = \left(\frac{41}{10}, -\frac{19}{6}, -7 \right),\\
\quad
&(b_1,b_2,b_3)_{\mathrm{MSSM}} = \left(\frac{33}{5}, 1, -3 \right),
\ea
\end{equation}
and
\begin{equation}
\frac{1}{\alpha_1(M_Z)} \simeq 59,\quad
\frac{1}{\alpha_2(M_Z)} \simeq 29.6,\quad
\frac{1}{\alpha_3(M_Z)} \simeq 8.5 \, .
\end{equation}

Now evaluate the running for two representative values of the supersymmetry breaking scale.

\begin{enumerate}
\item Lower values of $1/\alpha_i$: $M_{\text{SUSY}} = 10^{4}\,\text{GeV}$.

\begin{equation}
\ba
\ln\frac{M_{\text{SUSY}}}{M_Z} = \ln(10^2) \simeq 4.61\,,
\\
\ln\frac{\mu}{M_{\text{SUSY}}} = \ln(10^{10}) \simeq 23.03\,.
\ea
\end{equation}

\begin{align}
\frac{1}{\alpha_1} &\simeq 59 - 3.0 - 24.2 \simeq 31.8\,, \\
\frac{1}{\alpha_2} &\simeq 29.6 + 2.3 - 3.66 \simeq 28.2\,, \\
\frac{1}{\alpha_3} &\simeq 8.5 + 5.1 + 11.0 \simeq 24.6\,.
\end{align}

\item Higher values of $1/\alpha_i$: $M_{\text{SUSY}} = 10^{11}\,\text{GeV}$.

\begin{equation}
\ba
&\ln\frac{M_{\text{SUSY}}}{M_Z} = \ln(10^9) \simeq 20.72\,,
\\
&\ln\frac{\mu}{M_{\text{SUSY}}} = \ln(10^3) \simeq 6.91\,.
\ea
\end{equation}

\begin{align}
\frac{1}{\alpha_1} &\simeq 59 - 13.5 - 7.3 \simeq 38.2\,, \\
\frac{1}{\alpha_2} &\simeq 29.6 + 10.4 - 1.10 \simeq 38.9\,, \\
\frac{1}{\alpha_3} &\simeq 8.5 + 23.1 + 3.3 \simeq 34.9\,.
\end{align}
\end{enumerate}

Notice that $\alpha_1=\frac{5}{3}\alpha_Y$ with respect to the standard model $U(1)_Y$ normalization.

These results show that the gauge couplings approach each other at high energies while remaining non-universal, consistent with the non-GUT structure of the model. Here we neglect threshold effects associated with the top quark mass, since $m_t \sim M_Z$ and the corresponding logarithmic corrections are numerically subleading compared to the large running between $M_Z$ and $M_{\text{SUSY}}$. Besides, in our F-theory MSSM model, there may exist some additional vector-like pairs with zero net chirality, which contribute to the beta function in the high energy scale. The precise determination of the non-chiral matter spectrum requires delicate information on the $G_4$ flux, in terms of Deligne cohomology and root bundles~\cite{Bies:2014sra,Bies:2017fam,Bies:2020gvf,Bies:2021nje,Bies:2021xfh,Bies:2022wvj}. Thus this analysis can only give a lower bound on the coupling constants.

The $\theta_{QCD}$ and lower bounds of gauge coupling constants impose constraints on the moduli space. This allows us to exclude many geometry backgrounds.

\subsection{More than one axion}
\label{sec:more}

In most cases of F-theory compactification, $h^{1,1}(B_3)>1$, implying the existence of $h^{1,1}(B_3)$ axions (axion-like-particles), which may couple to the Standard Model QCD sector differently. We provide a brief discussion on this scenario. We consider the physics at the IR energy scale where the QCD axion acquires a potential term from the pion mass. The kinetic matrix in (\ref{axionkinetic}) and the mass matrix can be diagonalized simultaneously: one first performs a linear field redefinition that brings the kinetic terms to canonical form, and then an orthogonal rotation that diagonalizes the resulting mass matrix. Below we assume that the first step has already been carried out, i.e. the axions have been rescaled to canonically normalized fields $a\rightarrow a/f_a$, and all vectors in the following formulae are understood to be expressed in this canonically normalized basis. Assuming furthermore that the mass matrix from the UV potential is positive definite, the action becomes:
\be
\ba
S_{axion} = &\int  v^\top a \tr(F\wedge F)-\frac{1}{2}\int d^4x [\d_\mu a^\top \d^\mu a + a^\top M a]\,,\\
M=&m'^2v v^\top+\sum_\alpha m_\alpha^2 n_\alpha n_\alpha^\top\,,
\ea
\ee
where $m'$ is the mass contribution from the pion mass term, $v$ and $n_\alpha$ are $h^{1,1}\times 1$ vectors of real coefficients, and $m_\alpha$ are real scalars. $v$ captures the coupling of the multiple axions with the $SU(3)$ field strength $F$, which can be read off from (\ref{mult-axion-coupling}) or the F-theory derivation in Section~\ref{section3}. Throughout this section $F$ denotes the $SU(3)$ field strength, so that the decay constants $f_{a_i}$ defined below are the ones relevant for the coupling to gluons. The vectors $n_\alpha$ and masses $m_\alpha$ are determined by the UV potential (\ref{uvpo}). Here $a=(a_1, a_2, ..., a_{h^{1,1}})^\top$, where $a_i$ denotes the $i$-th axion. Let $\{b_i\}$ be the orthonormal eigenvector basis of $M$. After diagonalizing the mass matrix using $\{b_i\}$, the mass of the $i$-th axion and its coupling to instantons are given by:
\be
\ba
m_i&=\sqrt{m'^2(v\cdot b_i)^2+\sum_\alpha m_\alpha^2(n_\alpha\cdot b_i)^2}\,,\\
g_{i,agg}&= v \cdot b_i\propto \frac{1}{f_{a_i}} \\
\ea
\ee
Thus,
\be
\left|\frac{g_{i,agg}}{m_i}\right|= \left|\frac{v \cdot b_i}{\sqrt{m'^2(v\cdot b_i)^2+\sum_\alpha m_\alpha^2(n_\alpha\cdot b_i)^2}}\right|\leq \frac{1}{m'}\,.
\ee
This implies that all axions in our models, with $1/f_{a_i}$ defined through their coupling to $SU(3)$, lie below the QCD axion line in Fig.~\ref{m-f2}. If the mass matrix associated with the UV potential is not positive definite, it can generically possess at most one negative eigenvalue. The corresponding non-perturbative contribution is then unable to stabilize the associated axionic direction at its own minimum. Instead, this direction is stabilized predominantly by the IR potential, while the UV contribution shifts the vacuum away from the CP-conserving minimum, thereby inducing a non-zero QCD vacuum angle. Since the experimental constraint requires $\theta_{\mathrm{QCD}}<10^{-10}$, the magnitude of this negative UV contribution must be highly suppressed compared with that of the IR potential. Consequently, its effect on the second derivatives of the total potential is negligible, and it can be omitted in the computation of the axion mass spectrum. We stress that this conclusion holds for generic values of the phases $\phi_0$ and $\phi_\alpha$ in (\ref{uvpo}). If the phases are fine-tuned, the UV potential can lower the QCD axion mass without spoiling the solution to the strong CP problem. For instance, in the $h^{1,1}=1$ case, if the phase of the UV potential is tuned to exactly $\pi$, the UV contribution does not induce a non-zero CP violation angle, but it does reduce the mass of the QCD axion, so that the axion could lie above the QCD axion line. In other words, the statement that no axion lies above the QCD axion line follows only once fine-tuned phases are discarded and a small $\theta_{\rm QCD}$ is demanded. In the rest of this paper we focus on the generic case without fine-tuning.

\section{MSSM in F-theory and the axion coupling}
\label{section3}

For the MSSM construction in 4D F-theory, we use the class of model with the fiber polytope $F_{11}$ in \cite{Klevers:2014bqa,Cvetic:2015txa,Cvetic:2019gnh}, realizing the Standard Model gauge group $G_{SM}=SU(3)\times SU(2)\times U(1)_Y$ geometrically. The singular elliptic Calabi-Yau fourfold $X_4$ is described as a hypersurface equation 
\be
\label{F11}
s_1 u^3+ s_2 u^2 v + s_3uv^2 + s_5 u^2 w + s_6 uvw + s_9vw^2=0
\ee
where $[u:v:w]$ are projective coordinates of a $\mb{P}^2$. The $s_i$s are sections of line bundles
\be
\ba
[s_1]&=-3K_B-\mc{S}_7-\mc{S}_9\ ,\ [s_2]=-2K_B-\mc{S}_9\ ,
[s_3]=-K_B+\mc{S}_7-\mc{S}_9\ , \cr 
\ [s_5]&=-2K_B-\mc{S}_7\ ,
[s_6]=-K_B\ ,\ [s_9]=\mc{S}_9\,,
\ea
\label{divisor}
\ee
where $-K_B$ is the anticanonical divisor of the base threefold $B_3$. These line bundles are required to be effective, and the choices of $\mc{S}_7$, $\mc{S}_9$ specify the fibration.

The Weierstrass polynomials corresponding to (\ref{F11}) are
\be
\ba
f=&\frac{1}{2}s_3 s_9(-s_5 s_6+2s_1 s_9)-\frac{1}{48}(s_6^2-4s_2 s_9)^2\,,\cr
g=&\frac{1}{864}(216 s_3^2 s_5^2 s_9^2+36s_3 s_9(s_5 s_6-2s_1 s_9)(s_6^2-4s_2 s_9)\cr
&+(s_6^2-4s_2 s_9)^3)\,.
\ea
\ee

The discriminant locus reads
\be
\ba
\Delta&=\frac{1}{16}(s_3 s_5^3 s_6^3-s_2 s_5^2 s_6^4+s_1 s_5 s_6^5)s_3^2 s_9^3\cr
&+(\text{higher order terms})
\ea
\ee

In particular, the $SU(3)$ gauge group lives on the divisor $s_9=0$, corresponding to the divisor class $\mc{S}_9$. The $SU(2)$ gauge group lives on the divisor $s_3=0$, corresponding to the divisor class $-K_B+\mc{S}_7-\mc{S}_9$.

The singular Weierstrass model can be resolved by the blow-ups of 5D toric ambient space~\cite{Klevers:2014bqa}:
\be
\label{resolution}
(u,w;e_1)\ ,\ (u,v;e_2)\ ,\ (e_2,v;e_3)\ ,\ (u,e_1;e_4)\,.
\ee
The exceptional divisor corresponding to the Cartan subalgebra of the $SU(2)$ gauge group is $D^{SU(2)}_1=[e_1]$, and the ones corresponding to the Cartan subalgebra of the $SU(3)$ gauge group are $D^{SU(3)}_1=[e_2]$, $D^{SU(3)}_2=[u]$.

The Shioda map of the additional rational section $\hat{s}_1$ leading to a $U(1)_Y$ gauge group reads
\be
\sigma(\hat{s}_1)=S_1-S_0+[K_B]+\frac{1}{2}D^{SU(2)}_1+\frac{1}{3}(D^{SU(3)}_1+2D^{SU(3)}_2)\,.
\ee
The height pairing between $\hat{s}_1$ and itself is
\be
b_{11}=-\frac{3}{2}K_B-\frac{1}{2}\mc{S}_7-\frac{1}{6}\mc{S}_9\,,
\ee
leading to the coefficient of the kinetic term of $U(1)$:
\be
f_{11}=\frac{1}{2}\int_{B_3}J\wedge J\wedge b_{11}=\text{Vol}(b_{11})\,.
\ee
The gauge coupling for $SU(3)$, $SU(2)$ and $U(1)_Y$ factors are generally read off from the volume of divisors supporting the gauge group:
\be
\ba
g_3^2&=\frac{2\pi}{\text{Vol}(\mc{S}_9)}\,,\cr
g_2^2&=\frac{2\pi}{\text{Vol}(-K_B+\mc{S}_7-\mc{S}_9)}\,,\cr
g_Y^2&=\frac{\pi}{\text{Vol}(b_{11})}\,.
\ea
\ee

In the special case analyzed in \cite{Cvetic:2019gnh}, if we take $\mc{S}_7=\mc{S}_9=-K_B$, then both of the $SU(3)$ and $SU(2)$ gauge groups live on the anticanonical divisor class $-K_B$, and 
\be
b_{11}=-\frac{5}{6}K_B\,.
\ee

The gauge coupling of $SU(3)$ and $SU(2)$ gauge groups are
\be
g_3^2=g_2^2=\frac{2\pi}{\text{Vol}(-K_B)}
\ee
while the gauge coupling of the $U(1)_Y$ is
\be
g_Y^2=\frac{\pi}{\text{Vol}(-\frac{5}{6}K_B)}=\frac{3}{5}g_3^2\,,
\ee
i.e. $g_1^2=g_2^2=g_3^2$ in the GUT normalization $g_1=\sqrt{5/3}\,g_Y$, which resembles gauge coupling unification at the compactification scale. We stress that this relation is an accidental feature of the particular choice $\mc{S}_7=\mc{S}_9=-K_B$, for which all three gauge groups are supported on divisor classes proportional to $-K_B$; for generic choices of $\mc{S}_7$ and $\mc{S}_9$ the three gauge couplings are independent, and our analysis does not rely on unification. In particular, the one-loop running of section~\ref{sec:RG} with the minimal MSSM spectrum does not lead to universal couplings at $\mu\sim 10^{14}$ GeV. This is not a contradiction, since the values obtained in section~\ref{sec:RG} are only used as lower bounds on the gauge couplings at the compactification scale: additional vector-like matter, which is not fixed by our analysis, increases the couplings at high energies and may account for the difference between the extrapolated values and the geometric relation above.

Now let us discuss the coupling of axion to the Standard Model gauge fields from a UV perspective. We first discuss the axion coupling with the Mordell-Weil $U(1)_Y$. Recall that the axion comes from
\be
a_i=\int_{D_i}C_4\,,
\ee
where $D_i$ is the dual 4-cycle to $\omega_i\in H^{1,1}(B_3)$ on $B_3$.

Alternatively, we can expand
\be
C_4=\sum_i a_i \omega^{2,2}_i\,,
\ee
where $\omega^{2,2}_i$ is a $(2,2)$-form dual to a canonically chosen complex curve $\mc{C}_i$ on $B$, i.e.
\be
\label{Ci-Dj}
\mc{C}_i\cdot D_j=\delta_{ij}
\ee
with the basis of 4-cycles $D_j$. $\mc{C}_i$ can also be defined from the K\"{a}hler moduli parameter
\be
v^i=\int_{\mc{C}_i}J\,.
\ee
Since the Mordell-Weil $U(1)_Y$ is not supported on a single stack of 7-branes, its axion coupling cannot be read off directly from the 7-brane Chern-Simons action (\ref{axion_ac}) as in section~\ref{sec:EFF}. Instead, one expects the axion coupling term to the Mordell-Weil $U(1)_Y$ to take the form 
\be
\sim\int c_{i YY}a_i F_Y\wedge F_Y\,,
\ee
\be
c_{i YY}\propto(b_{11}\cdot \mc{C}_i)_{B_3}\,,
\ee
where $(b_{11}\cdot \mc{C}_i)_{B_3}$ denotes the intersection number between the divisor class $b_{11}$ and the curve $\mc{C}_i$ on $B_3$, which is a (rational) number. In the remainder of this section we derive this coefficient, as well as the couplings $c_{iGG}$ to the non-abelian gauge groups, from the dual M-theory picture, and show that the latter agree with the IIB result (\ref{IIB-axion-coup}).

We now derive the axion coupling from the M-theory dual picture. We start from the M-theory topological action
\be
\label{Mtop}
S_{top}=-\int_{\mc{M}_{11}}\frac{1}{6}C_3\wedge G_4\wedge G_4
\ee
and expand 
\be
\ba
C_3&=A_Y\wedge [\sigma(\hat{s}_1)]+\dots\cr
G_4&=F_Y\wedge [\sigma(\hat{s}_1)]+\sum_{i=1}^{h^{1,1}(B_3)}g_i\wedge \hat{\omega}^{2,2}_i\,.
\ea
\ee
$\hat{\omega}^{2,2}_i$ is the $(2,2)$-form in the elliptic CY4 that is dual to the 4-cycle $Z\cdot \hat{D}_i$, where $Z$ is the zero section of elliptic fibration, and $\hat{D}_i$ is the vertical divisor in the CY4 corresponding to $D_i$ on $B_3$. In the expansion, $g_i$ is a closed 0-form, which can be thought as the field strength of the ``$(-1)$-form'' field after the dimensional reduction of the 0-form axion $\theta_i$ to 3D.

More explicitly, we can make the following argument following the case of 6D F-theory~\cite{Grimm:2013oga}. In such case, the 6D $B\wedge F\wedge F$ term arises from the uplift of the 5D $A'\wedge F\wedge F$ term, where the 5D 1-form gauge field $A'$ is the dimensional reduction of the 6D self-dual 2-form $B$. One can simply check that $B$ and $A'$ have the same number of on-shell degrees of freedom, which is 3.

Now in the case of 4D F-theory, one should expect that the 4D axion coupling term $\theta_i F\wedge F$ similarly arises from the coupling with a ``$(-1)$-form'' gauge field, $\theta_{-1,i}\wedge F\wedge F\sim g_i A\wedge F$. Nonetheless such $\theta_{-1,i}$, which is EM dual to a massless 2-form $B_{2,i}^{\rm 3D}$ in 3D, has 0 on-shell degree of freedom. 

Alternatively, one can start from the 2-form $B_{2,i}$ EM dual to the axion $a_i$ in 4D. After dimensional reduction to 3D, $B_{2,i}$ is reduced to $B_{2,i}^{\rm 3D}$ and a 1-form gauge field $A_i^{\rm 3D}$. The $A_i^{\rm 3D}$ contains the 1 on-shell degree of freedom from the 4D axion, which is dual to a 3D scalar $a_{i,\text{3D}}$. In M-theory the 3D axion $a_{i,\text{3D}}$ arises from
\be
a_{i,\text{3D}}=\int_{\widehat{D}_i}C_6\,,
\ee
and its dual $A_i^{\rm 3D}$ arises from the expansion
\be
C_3=A_i^{\rm 3D}\wedge\hat{\omega}^{1,1}_i+\dots\,,
\ee
where $\hat{\omega}^{1,1}_i$ is dual to the vertical divisor $\hat{D}_i$ fibered over $D_i$ in the resolved CY4. 

The 3D axion $a_{i,\text{3D}}$ appears in the M-theory superpotential
\be
W=W_0+\sum_i A_i e^{-(V_{\hat{D}_i}+ia_{i,\text{3D}})}\,,
\ee
which can be naturally uplifted to the 4D axion superpotential (\ref{W-total}).

Thus after reduction on the elliptic CY4 $Y_4$, we obtain the coupling term
\be
-\frac{1}{2}\int_{Y_4}[\sigma(\hat{s}_1)]\wedge [\sigma(\hat{s}_1)]\wedge \hat{\omega}^{2,2}_i\int_{\mb{R}^{2,1}} g_i A_Y\wedge F_Y\,.
\ee
After uplifting to 4D, we replace $g_i\rightarrow \mathrm{d}a_i$ and give the axion coupling term in 4D
\be
\ba
-\int_{\mb{R}^{3,1}} c_{i YY}\mathrm{d}a_i\wedge A_Y\wedge F_Y\
=\int_{\mb{R}^{3,1}} c_{i YY}a_i F_Y\wedge F_Y\,.
\ea
\ee

Note that in the F-theory limit, we have 
\be
\ba
&\frac{1}{2}\int_{Y_4}[\sigma(\hat{s}_1)]\wedge [\sigma(\hat{s}_1)]\wedge \hat{\omega}^{2,2}_i =\frac{1}{2}\int_{B_3}[\pi_*(\sigma(\hat{s}_1)\cdot\sigma(\hat{s}_1))]\wedge \omega^{2,2}_i&=c_{i YY}\,.
\ea
\ee
\be
c_{i YY}=\frac{1}{2}(b_{11}\cdot \mc{C}_i)_{B_3}\,.
\ee

Then we compare the axion coupling $c_{i YY}$ with the axion coupling $c_{i GG}$ from the M-theory description, where $G$ is a non-abelian group supported on the divisor $D_G\subset B$. Denoting the exceptional divisors by $E_I$ $(I=1,\dots,\text{rk}(G))$, the intersection numbers are
\be
E_I\cdot E_J\cdot Z\cdot \hat{D}_i=-\mathfrak{C}_{IJ}(D_G\cdot \mc{C}_i)_{B_3}\,.
\ee
$\mathfrak{C}_{IJ}$ is the Cartan matrix of $G$.
For $c_{i GG}$, we only need to consider an abelian subgroup $U(1)_I\subset G$, generated by $E_I$. Expanding 
\be
\ba
C_3&=A_I\wedge[E_I]+\dots\cr
G_4&=F_I\wedge[E_I]+\sum_{i=1}^{h^{1,1}(B_3)}g_i\wedge \hat{\omega}^{2,2}_i\,.
\ea
\ee
From (\ref{Mtop}) we get the topological term
\be
\ba
&-\frac{1}{2}\int_{Y_4}[E_I]\wedge[E_I]\wedge \hat{\omega}^{2,2}_i\int_{\mb{R}^{2,1}}g_i A_I\wedge F_I =
-(D_G\cdot \mc{C}_i)_{B_3}\int_{\mb{R}^{2,1}}g_i A_I\wedge F_I\,.
\ea
\ee
Similar to before, after uplifting to 4d, we get the axion coupling term
\be
(D_G\cdot \mc{C}_i)_{B_3}\int_{\mb{R}^{3,1}}a_i F_I\wedge F_I\,.
\ee
Using standard Killing form $\tr(t^a t^b)=\frac{1}{2}\delta^{ab}$, it can be rewritten as
\be
\int_{\mb{R}^{3,1}}c_{i GG}a_i\tr(F\wedge F)\,,
\ee
with coefficient
\be
\label{ciGG}
c_{i GG}=2(D_G\cdot \mc{C}_i)_{B_3}\,.
\ee
When $\mc{S}_7=\mc{S}_9=-K_B$, we have the relation
\be
c_{i GG}=\frac{24}{5}c_{i YY}\,.
\ee
To match the coefficients $c_{iGG}$ with (\ref{IIB-axion-coup}), note that the curve $\mc{C}_i$ is exactly defined through (\ref{Ci-Dj}), i.e. if $D_G=\sum_i n_i D_i$, we have $(D_G\cdot \mc{C}_i)_{B_3}=n_i$, hence $c_{iGG}=2n_i$ in (\ref{IIB-axion-coup}), and the axion coupling terms derived with the two methods only differ by a normalization coefficient $8\pi$.

\section{Results for different base geometries}
\label{sec:results}

In this section, we compare the constraints from $\theta_{\rm QCD}$ and the gauge couplings on the K\"ahler moduli space for different base geometries. Our goal is not only to determine whether there exist points in the stretched K\"ahler cone satisfying
\[
\theta_{\rm QCD}<10^{-10},
\]
but also to understand whether the corresponding gauge couplings can be sufficiently large from the phenomenological point of view.

A basic difficulty is that the supersymmetry breaking scale is not known. Since the gauge couplings run logarithmically, a lower supersymmetry breaking scale generally leads to larger gauge couplings at the UV scale relevant to our geometric construction. In addition, if one allows extra vector-like pairs of fermions, then the gauge couplings are also enhanced under the RG flow. Therefore, within our present analysis, we cannot assign a unique value to the gauge couplings at the compactification scale. Instead, what can be robustly extracted is a conservative lower bound on each gauge coupling, obtained in the minimal case without additional fermions. Any extra fermion or lower supersymmetry breaking scale would only increase the gauge couplings.

Motivated by this, we divide the models into three classes, denoted by ``\texttt{Y}'', ``\texttt{O}'', and ``\texttt{N}''. A model is said to be of type ``\texttt{Y}'' if, for any supersymmetry breaking scale within the range we regard as physically plausible, one can find points in the allowed moduli space such that $\theta_{\rm QCD}<10^{-10}$ and all three gauge couplings are simultaneously above their respective lower bounds. In this sense, the ``\texttt{Y}'' models are robustly viable against the uncertainty in the RG running, if no additional non-chiral matter fields are introduced. On the other hand, a model is said to be of type ``\texttt{N}'' if, for every point in the allowed moduli space satisfying $\theta_{\rm QCD}<10^{-10}$, at least one gauge coupling is below its corresponding lower bound. Such models are therefore robustly excluded in our setup. The remaining models are classified as type ``\texttt{O}'': they are not definitively excluded, but their viability depends on details of the supersymmetry breaking scale. To make the classification more conservative, we slightly enlarge the parameter region assigned to the class ``\texttt{O}''. The detailed classification criteria are given in Appendix~\ref{app:data-files}; the bounds used there are slightly shifted with respect to the values obtained in section~\ref{sec:RG}.

With this classification in hand, we now study the explicit base geometries $\mb{P}^3$, $\mb{P}^1\times \mb{P}^2$, $\tilde{\mb{F}}_3$, and $\mb{P}^1\times \mb{P}^1\times \mb{P}^1$ in turn.

We review the subsets of F-theory base landscape, potentially leading to 4D standard-model-like models.

In particular, we use the ``quadrillion'' ensemble of models in \cite{Cvetic:2019gnh}, with an exact Standard Model particle spectrum after an appropriate choice of $G_4$ flux.

In this special subclass, the base threefold $B_3$ is given by one of the 4319 3D reflexive polytopes~\cite{Kreuzer:1998vb}. Different fibrations over $B_3$ are characterized by the line bundles $\mc{S}_7$ and $\mc{S}_9$. In the simplest case with $\mc{S}_7=\mc{S}_9=-K_B$, the $SU(2)$ and $SU(3)$ are over the divisor $-K_B$, and the height pairing $b_{11}=-\frac{5}{6}K_B$. We will by default use this criterion in the following discussions while the other choices for $\mc{S}_7$ and $\mc{S}_9$ will be specified otherwise. In this setting, the QCD gauge coupling $\alpha_3$ sets the strictest bounds, and we will only mention $\alpha_3$ and the axion coupling with the QCD sector in this section. Note that, since the $SU(2)$ gauge group is supported on the same divisor $-K_B$ as $SU(3)$ and $c_{iGG}=\frac{24}{5}c_{iYY}$ for all $i$ when $\mc{S}_7=\mc{S}_9=-K_B$, the couplings of each axion mass eigenstate to $SU(2)$ and $U(1)_Y$ follow from its $SU(3)$ coupling quoted below by the universal factors $1$ and $5/24$, respectively. In short, the allowed models with $\alpha_3\geq 1/23$ are labeled as ``\texttt{Y}'', the ones with $1/23>\alpha_3\geq 1/37$ are labeled with ``\texttt{O}'', and the ones with $\alpha_3<1/37$ are labeled as ``\texttt{N}''.

\subsection{$\mb{P}^3$}
\label{p3ax}
The simplest example with $h^{1,1}(B_3)=1$ is $B_3=\mb{P}^3$, with a single divisor class $D_1=H$ that is the hyperplane class, and the triple intersection number $H^3=1$ on the base. Hence the canonical basis for divisor and curves dual to each other are
\be
D_1=H\ ,\ \mc{C}_1=H\cdot H\,.
\ee
The anticanonical divisor is $-K_{\mb{P}^3}=4H$. The K\"ahler modulus is defined as
\be
v^1=\int_{H\cdot H}J
\ee
or equivalently 
\be
J=v^1\omega_1=v^1[H]\,.
\ee
Since $B_3$ is toric, its K\"ahler cone is known exactly; for $\mb{P}^3$ it is simply $v^1>0$, and the 1-stretched K\"ahler cone condition reads $v^1>1$. We will state the K\"ahler cone explicitly at the outset for each of the bases considered below.
The volume of $B_3$ is
\be
\text{Vol}(B_3)=\frac{1}{6}(v^1)^3\,,
\ee
while the volume of $-K_{\mb{P}^3}=4H$ is
\be
\text{Vol}(4H)=\frac{1}{2}4H\cdot (v^1 H)\cdot (v^1 H)=2(v^1)^2\,.
\ee
Thus
\be
g_3^2=g_2^2=\pi/(v^1)^2\,,
\ee
\be
g_Y^2=\frac{\pi}{\text{Vol}(\frac{10}{3}H)}=\frac{3}{5}\pi/(v^1)^2\,.
\ee
We can compute $c_{1 YY}=\frac{1}{2} (b_{11}\cdot \mc{C}_1)_{B_3}=5/3$.

The coupling with non-abelian gauge groups is\footnote{As explained before, it would lead to the axion coupling term $\frac{1}{4\pi}\int 4a\tr(F_3\wedge F_3)$ in the normalization of (\ref{IIB-axion-coup}).}
\be
c_{1 GG}=2((-K_B)\cdot \mc{C}_1)_{B_3}=8\,.
\ee
We construct an MSSM in $\mb{P}^3$ with gauge group $\mathrm{SU}(3)\times\mathrm{SU}(2)\times\mathrm{U}(1)$. Since we are interested in $\theta_{\text{QCD}}$, we focus only on the $\mathrm{SU}(3)$ gauge group in the following discussion. Geometrically $[H]$ is not a rigid divisor on $B_3$, but to obtain a non-zero instanton superpotential, we assume that $[H]$ is rigidified by an appropriate choice of flux\footnote{The detailed analysis of rigidification on the specific CY4 would be left in a future work.}. After simplifying Eqs.~(\ref{axionkinetic}) and (\ref{uvpo}), we can write down the UV action of the axion and the $\mathrm{SU}(3)$ gauge field:
\be
\ba
S=& -\int \mathrm{d}^4 x\sqrt{-g}\frac{1}{2} \frac{2(v^1)^2}{2\pi}\tr(F_{\mu\nu,3}F^{\mu\nu}_3)+\frac{1}{4\pi}\int 4 a_1 \tr(F_{3}\wedge F_{3})\\
&-\frac{1}{2}\int\mathrm{d}^4x\sqrt{-g}\frac{6}{(v^1)^4}\d_\mu a_1\d^\mu a_1-\int \mathrm{d}^4 x \frac{144\pi C}{(v^1)^4}e^{-\pi (v^1)^2}\cos(2\pi a_1+\phi_0-\phi_1),\\
C=&\left|\frac{A_1 W_0}{\int_{\hat{X}_4}\Omega\wedge\bar{\Omega}-\sum_S \int_S\Tr(\phi'\wedge\bar{\phi}')}\right|\,.
\ea
\ee

The UV scalar potential is \footnote{Of course contributions from E3 brane wrapping other 4-cycles $aH$ with $a>1$ are also expected, but due to the exponential suppression of the $e^{-\pi (v^1)^2}$ factor, these additional terms will be ignored. Similar arguments apply to other examples in the later Sections.}:
\be
\mc{V}_{\text{UV}}=\frac{144\pi C}{(v^1)^4}e^{-\pi (v^1)^2}\cos(2\pi a_1+\phi_0-\phi_1)\,.
\ee

From Eqs.~(\ref{QCDaxion}) and (\ref{string_angle}), we obtain:
\be
\ba
f_{a}&=\frac{\sqrt{6}}{8\pi (v^1)^2}\approx 1.6(v^1)^{-7/2}\times10^{17}\mathrm{GeV}\,,\\
\theta_{QCD}&=\langle 2\pi \times 4a_1\rangle\,.
\ea
\ee
Note that $f_a$ is invariant under RG flow. Considering the IR contribution to the axion potential (\ref{uvpion}):
\be
\ba
\mc{V}_{\text{IR}}&=\frac{2 m_{\pi}^2 f_\pi^2m_um_d}{(m_u+m_d)^2}\sin^2(4\pi a_1)\approx \frac{1}{2} 1.9\times 10^{-77}(v^1)^6 \sin^2(4\pi a_1)\,.
\ea
\ee
One can obtain $\theta_{\text{QCD}}$ by solving for the minimum of $\mc{V}_{\text{UV}}+\mc{V}_{\text{IR}}$. Since $|\theta_{\text{QCD}}|<10^{-10}$, we have $\sin(4\pi a_1) \approx4\pi a_1$ and $\cos(2\pi a_1+\phi_0-\phi_1)\approx\cos(\phi_0 -\phi_1)$. We obtain:
\be
\ba
\label{thetaQCD-P3}
|\theta_{QCD}| &\approx 2.4\times 10^{79} \frac{C}{(v^1)^{10}}e^{-\pi (v^1)^2}\lvert \sin(\phi_0-\phi_1)\rvert\approx 2.4 \times 10^{79} \frac{1}{(v^1)^{10}}e^{-\pi (v^1)^2},
\ea
\ee
and the UV contribution to the axion mass\footnote{Note that $m_{a'}$ can be much smaller than the axion mass generated by QCD effects in the IR, hence it cannot be sensitively measured by the current experiments.}
\be
\ba
m_{\text{UV}}^2=&96 \pi^3 e^{-\pi (v^1)^2}C\lvert\cos (\phi_0-\phi_1)\rvert[1.7(v^1)^{-3/2}\times10^{18} \mathrm{GeV}]^2.
\ea
\label{p3cpva}
\ee
The IR contribution to axion mass is
\be
m_{\text{IR}}^2\approx [6\times 10^{-6}\text{eV}\left(\frac{10^{12}\text{GeV}}{f_a}\right)]^2\approx [4\times10^{-11}(v^1)^{7/2}\text{eV}]^2.
\ee

Notice that the dominant factor in (\ref{thetaQCD-P3}) is $e^{-\pi (v^1)^2}$, and the lower bound of $v^1$ is insensitive to the value of $|C\sin (\phi_0-\phi_1)|$. For the latter factor, we do not impose particular constraints on $(\phi_0-\phi_1)$. We assume that the geometric model lies in a generic point of the moduli space, which is reasonable for our F-theory setup. Estimating $|C\sin (\phi_0-\phi_1)|\approx 1$ and using the experimental bound $|\langle a'/f_a\rangle|<10^{-10}$, we obtain the lower bound on the K\"ahler parameter: 
\be
\label{v1-bound-P3}
v^1\gtrsim7.7\,.
\ee
There may be some IR effects along the RG flow that modify the axion potential derived from $W_{np}$ in the UV. However, such effects can only change the prefactor of $\mc{V}_{UV}$, and not the exponential suppression factor $e^{-\pi(v^1)^2}$ that dominates (\ref{thetaQCD-P3}). As can be seen from Figs.~\ref{P1P21}-\ref{F34}, changing the numerical coefficient in front of $\mc{V}_{UV}$ by three orders of magnitude does not affect the exclusion curves significantly. Hence our estimate of the lower bound on $v^1$ remains valid. 

There is another bound on $v^1$. From Eq.~(\ref{guagefieldnab}), we know $g^2=\frac{4\pi}{2 V_{\text{4cycle}}}=\frac{4\pi}{4(v^1)^2}$. Thus we obtain a relation between the K\"ahler parameter $v^1$ and the $\alpha_3$ measured in the UV (below the KK scale):
\be
v^1=\sqrt{\frac{1}{4\alpha_3}}\,.
\ee
If we choose $\alpha_3=1/37$(the lower bound of ``\texttt{O}''), then $v^1=3.0$, contradicts the bound (\ref{v1-bound-P3}) derived from CP violation angle. This model is ``\texttt{N}'' in our classification. This implies that our assumption that $[H]$ is rigidified is non-valid, and $[H]$ must be a non-rigid divisor. We can change the rigidified divisor to find a model classified as ``\texttt{O}'' or ``\texttt{Y}''. Assuming the divisor $[nH]$ is rigidified, we obtain

\be
\ba
|\theta_{QCD}|&\approx 9.5\times10^{79} \frac{n^2}{4 (v^1)^{10}}e^{-n\pi (v^1)^2},\\
\alpha_3&=\frac{1}{4 (v^1)^2}\,,\\
f_a&=\frac{\sqrt{6}}{8\pi (v^1)^2}\approx 1.6(v^1)^{-7/2}\times 10^{17} \mathrm{GeV}\,.
\ea
\ee

We can acquire the lower bound of $v^1$ for $\theta_{QCD}\leq 10^{-10}$, listed in Table~\ref{table1}.
\begin{table}[h]
    \centering
    \begin{tabular}{|c|cccccc|}
    \hline
       $n$  &1 &2&3&4&5&6 \\
       \hline
        $v^1$ & 7.7&5.5&4.5&3.9&3.5&3.2\\
    \hline
    \hline
       $n$  &7 &8&9&10&11&12 \\
       \hline
        $v^1$ & 3.0&2.8&2.7&2.5&2.4&2.3\\
    \hline
    \end{tabular}
    \caption{The lower bound of $v^1$ from the condition $\theta_{\text{QCD}}\leq 10^{-10}$ for rigidified divisor $[nH]$.}
    \label{table1}
\end{table}

Now we can find the bounds for ``\texttt{Y}'',``\texttt{O}'',``\texttt{N}'' are $v^1\leq 2.4$(more precisely, 2.398, so n=11 is ``\texttt{O}''), $2.4<v^1\leq 3.0$, $v^1>3.0$ respectively. Thus, for $[nH]$ being the rigidified divisor, $n\leq 6$ is ``\texttt{N}'',  $6<n\leq 11$ is ``\texttt{O}'', and $n>11$ is ``\texttt{Y}''. 

Besides, we can find that the mass contributed by the UV potential is
\be
\ba
|m_{\text{UV}}^2| &\approx |\theta_{QCD}|\, 4n(v^1)^7 \times 10^{-22}(\text{eV})^2\approx|\theta_{QCD}|\frac{n}{4} m_{\text{IR}}^2\ll m_{\text{IR}}^2\,,
\ea
\ee
which is far less than the mass contributed by the pion mass term. For the axion decay constant and mass, we choose a representative point of $v^1=2.5$, and get
\be
\ba
&f_a\approx6.5\times10^{15}\text{GeV}\,,\\
&m_a\approx 9.9\times 10^{-10}\text{eV}\,.
\ea
\ee

\subsection{$\mb{P}^1\times\mb{P}^2$}
\label{P1P2ax}
Now we consider the base $B_3=\mb{P}^1\times\mb{P}^2$ with $h^{1,1}(B_3)=2$. There are two divisor classes $D_1=S\cong \mb{P}^2$ and $D_2=H\cong \mb{P}^1\times\mb{P}^1$, neither of which is rigid without the inclusion of flux, with the triple intersection numbers on $B_3$:
\be
H^3=0\ ,\ S\cdot H^2=1\ ,\ S^2\cdot H=0\ ,\ S^3=0\,.
\ee

The canonical basis for divisor and curves are
\be
D_1=S\ ,\ D_2=H\ ,\ \mc{C}_1=H\cdot H\ ,\ \mc{C}_2=S\cdot H\,,
\ee
such that $D_i\cdot \mc{C}_j=\delta_{ij}$.

The K\"{a}hler form $J$ is
\be
J=v^1[S]+v^2[H]\,.
\ee
The volume of $B_3$ is
\be
\text{Vol}(B_3)=\frac{1}{2}v^1(v^2)^2\,.
\ee
The volume of $D_1$ and $D_2$ are
\be
\text{Vol}(D_1)=\frac{1}{2}(v^2)^2\ ,\ \text{Vol}(D_2)=v^1 v^2\,.
\ee
The anticanonical class $-K_{\mb{P}^1\times\mb{P}^2}=3H+2S$, with volume
\be
\ba
\label{VolK-P1P2}
\text{Vol}(3H+2S)&=\frac{1}{2}(3H+2S)(v^1 S+v^2 H)(v^1 S+v^2 H) =v^2(3v^1+v^2)\,.
\ea
\ee

The K\"ahler cone of $B_3$ is
\be
\{(v^1,v^2)\,|\,v^1>0,\ v^2>0\}\,,
\ee
so that the 1-stretched K\"ahler cone condition reads $v^1>1$, $v^2>1$.

The axion coupling with Mordell-Weil $U(1)$
\be
\ba
c_{1YY}&=\frac{1}{2}(-\frac{5}{6}K_B\cdot\mc{C}_1)=\frac{5}{6}\,,\cr
c_{2YY}&=\frac{1}{2}(-\frac{5}{6}K_B\cdot\mc{C}_2)=\frac{5}{4}\,.
\ea
\ee
The axion couplings with non-abelian gauge groups are
\be
\ba
c_{1GG}&=2(-K_B\cdot\mc{C}_1)=4\,,\cr
c_{2GG}&=2(-K_B\cdot\mc{C}_2)=6\,.
\ea
\ee 
We continue to focus on the axion coupling with the $\mathrm{SU}(3)$ gauge group supported on $-K_B=3H+2S$. We first consider a simplified case in which only the divisor classes $[H]$ and $[S]$ contribute to the superpotential, i.e. both $[H]$ and $[S]$ are assumed to be rigidified by flux. After simplifying Eqs.~(\ref{axionkinetic}), (\ref{uvpo}) and (\ref{uvpion}), the actions and scalar potential are given by:
\be
\ba
S_{\mathrm{SU(3)}}=& -2\pi \int \mathrm{d}^4 x (2 V_1+3V_2) \sqrt{-g}\frac{1}{8\pi^2} \tr(F_{\mu\nu,3} F^{\mu\nu}_3) +\frac{1}{4\pi}\int ( 2 a_1+3 a_2)\tr (F_3\wedge F_3)\,,\\
S_{\text{axion},k}=&-\frac{1}{2}\int \mathrm{d}^4 x \left[\frac{2}{(v^2)^4} \d_\mu a_1\d^\mu a_1+\frac{1}{(v^1)^2(v^2)^2}\d_\mu a_2 \d^\mu a_2\right],\\
\mc{V}_{\text{IR}}=&\frac{1}{2} 1.7\times 10^{-76}(v^1)^2(v^2)^4 \sin^2[\pi (2 a_1+3 a_2)]\,,\\
\mc{V}_{\text{UV}}=&\frac{16\pi C_1 }{(v^1)^2(v^2)^2}e^{-\pi(v^2)^2}\cos \psi_1 +\frac{32\pi C_2}{v^1(v^2)^3}e^{-2\pi v^1 v^2}\cos \psi_2\,,
\ea
\ee
where
\be
\ba
\psi_1&=2\pi a_1+\phi_0-\phi_1\,,\quad \psi_2=2\pi a_2+\phi_0-\phi_2\,,\\
C_1&=\left|\frac{W_0A_1}{\int_{\hat{X}_4}\Omega\wedge\bar{\Omega}-\sum_S \int_S\Tr(\phi'\wedge\bar{\phi}')}\right|,\\
C_2&=\left|\frac{W_0A_2}{\int_{\hat{X}_4}\Omega\wedge\bar{\Omega}-\sum_S \int_S\Tr(\phi'\wedge\bar{\phi}')}\right|.
\ea
\ee
We can solve for the minimum of the potential $V$ and obtain:
\be
\ba
\theta_{QCD}&=\langle 2\pi (2 a_1+3 a_2)\rangle \approx \frac{2.4\times10^{78}}{(v^1)^3(v^2)^6}\frac{C_1}{4v^1}e^{-\pi(v^2)^2}\sin\langle \psi_1\rangle=\frac{2.4\times10^{78}}{(v^1)^3(v^2)^6} \frac{C_2}{3v^2}e^{-2\pi v^1 v^2}\sin\langle\psi_2\rangle\,.
\ea
\label{P1P2cp}
\ee
The values of $\phi_0$, $\phi_1$, and $\phi_2$ remain undetermined. We assume they are random $\mathcal{O}(1)$ parameters. Thus, it is highly improbable for $\theta_{\text{QCD}}$, $\sin\langle \psi_1\rangle$, and $\sin\langle \psi_2\rangle$ to be simultaneously far less than 1. Based on the last equation of (\ref{P1P2cp}), we assert that for $\sin\langle \psi_i\rangle$ with a small coefficient, its value should not be significantly less than unity. We estimate:
\be
\theta_{QCD}\approx \frac{2.4\times10^{78}}{(v^1)^3(v^2)^6}\mathrm{min}\left(\frac{C_1}{4v^1}e^{-\pi(v^2)^2},\frac{C_2}{3v^2}e^{-2\pi v^1 v^2}\right)\,.
\ee
Furthermore, with the volume formula (\ref{VolK-P1P2}) of $-K_B$ supporting the $SU(3)$ gauge group, we obtain the QCD coupling constant in the UV:
\be
\alpha_3=\frac{1}{2(v^2)^2+6v^1v^2}\,.
\ee

\begin{figure}[h]
    \centering
    \includegraphics[width=0.8\linewidth]{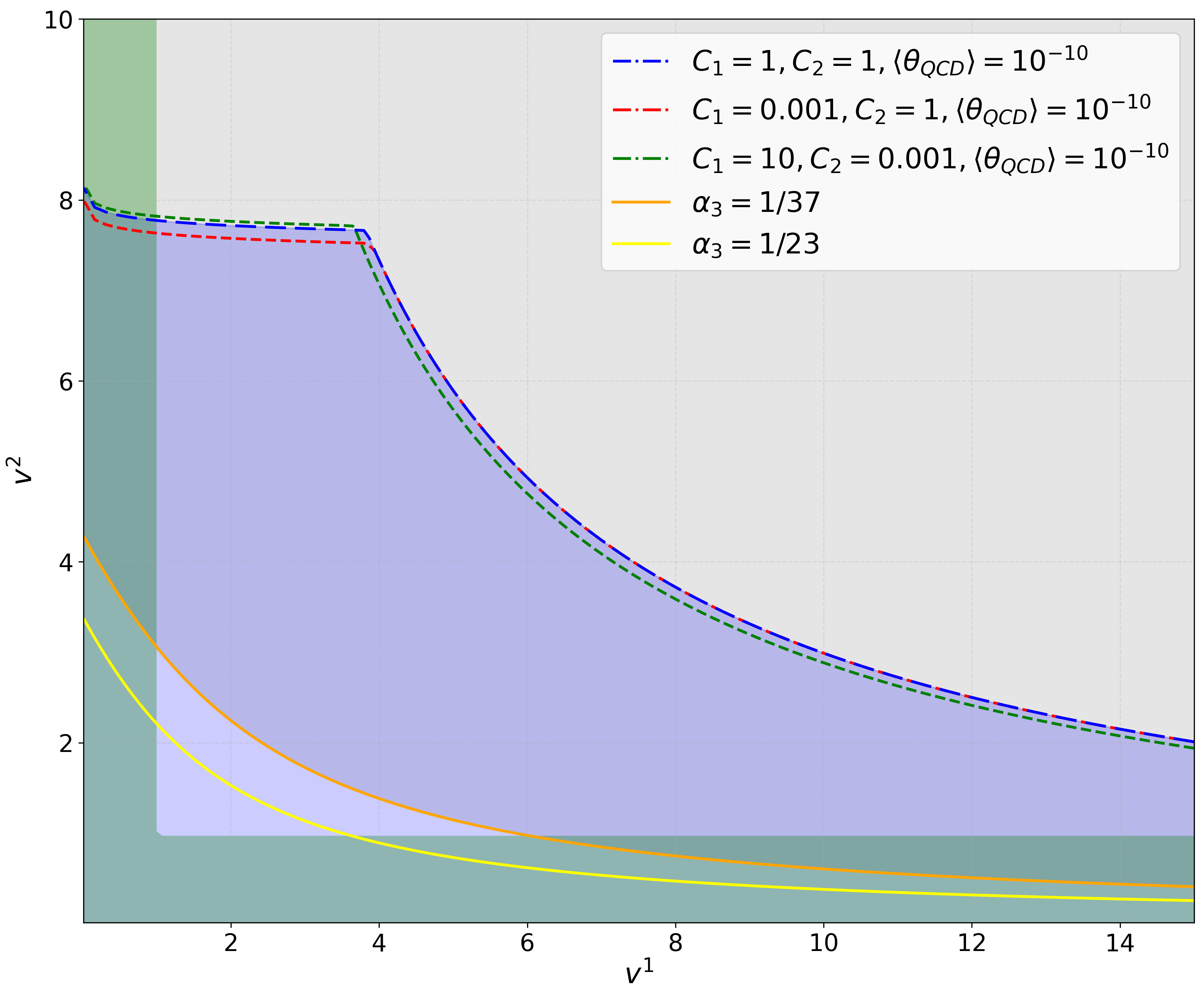}
    \caption{Geometric model over the base threefold $B_3=\mb{P}^1\times\mb{P}^2$. $\mc{S}_7=\mc{S}_9=-K_B$ with the set of rigid (or rigidified) divisors: $\{[S],[H]\}$. The horizontal axis $v^1$ and vertical axis $v^2$ are the coefficients of the two-forms in the K\"ahler form dual to $[S]$ and $[H]$, respectively. The dashed lines represent $\theta_{\text{QCD}}=10^{-10}$ for different values of $C_1$ and $C_2$. The solid orange line represents $\alpha_3=1/37$, and the solid yellow line represents $\alpha_3=1/23$. The blue region is excluded by $\theta_{\text{QCD}}\leq10^{-10}$, the gray region is excluded by $\alpha_3\geq 1/37$, and the green region is excluded by the stretched K\"ahler cone condition $v^1,v^2>1$. The dashed lines are closely clustered, indicating that our result is insensitive to the values of $C_1$ and $C_2$. From this figure, we observe that the entire moduli space is excluded. Consequently, this model is excluded, and is classified as ``\texttt{N}''.}
    \label{P1P21}
\end{figure}

These two equations impose constraints on $v^1$ and $v^2$. For clarity, all three constraints, namely $\theta_{QCD}<10^{-10}$, the gauge coupling bound $\alpha_3>1/37$ and the 1-stretched K\"ahler cone condition $v^1,v^2>1$, are shown in the same plot in Fig.~\ref{P1P21}. Additionally, to address the undetermined parameters $C_1$ and $C_2$, we vary their values by a few orders of magnitude to demonstrate the insensitivity of our model.

From Fig.~\ref{P1P21}, we find that the result is insensitive to the values of $C_1$ and $C_2$. Even though the UV potential may receive corrections as the energy scale decreases, the result will not change significantly. Fig.~\ref{P1P21} shows that the curve $\alpha_3=1/37$ does not enter the allowed region. In this model, we cannot find $(v^1,v^2)$ that satisfies $\theta_{\text{QCD}}\leq 10^{-10}$ and $\alpha_3\geq 1/37$ simultaneously. Therefore, we can exclude this geometric background; $[H]$ and $[S]$ cannot be rigid divisors at the same time. We require specific fluxes to rigidify or de-rigidify divisors, and thus need to consider other cases. If we have only one rigid divisor, aside from the elements of $[n(2S+3H)]$, we can obtain a CP violation angle of zero with $(v^1,v^2)$ at any locus in the K\"ahler cone.

\begin{figure}[h]
    \centering
    \includegraphics[width=0.8\linewidth]{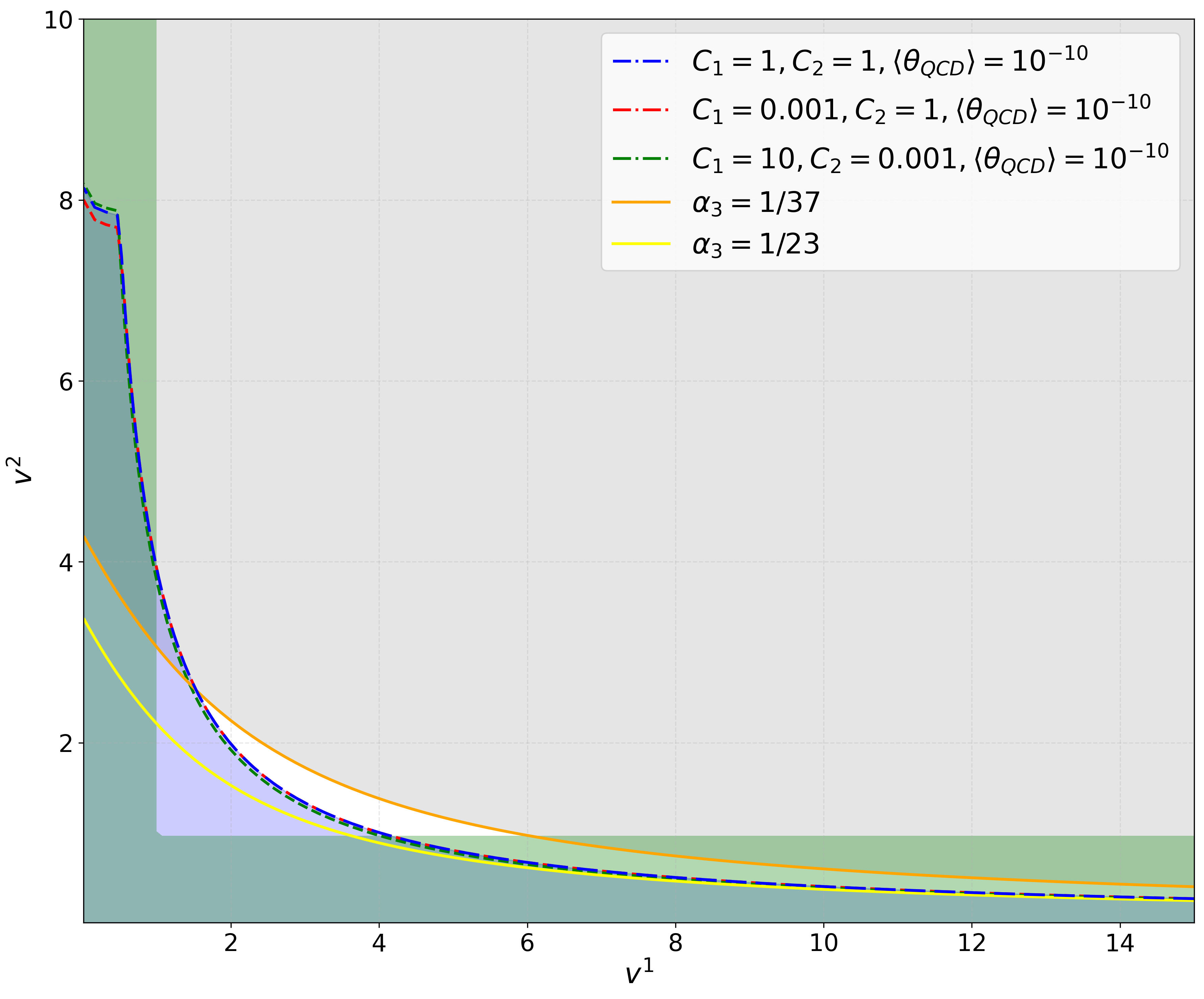}
    \caption{Geometric model over the base threefold $B_3=\mb{P}^1\times\mb{P}^2$. $\mc{S}_7=\mc{S}_9=-K_B$ with the set of rigid (or rigidified) divisors $\{[S],[8H]\}$. The horizontal axis $v^1$ and vertical axis $v^2$ are the coefficients of the two-forms in the K\"ahler form dual to $[S]$ and $[H]$, respectively. The dashed lines represent $\theta_{\text{QCD}}=10^{-10}$ for different values of $C_1$ and $C_2$. The solid orange line represents $\alpha_3=1/37$, and the solid yellow line represents $\alpha_3=1/23$. The blue region is excluded by $\theta_{\text{QCD}}\leq10^{-10}$, the gray region is excluded by $\alpha_3\geq 1/37$, and the green region is excluded by the stretched K\"ahler cone condition $v^1,v^2>1$. The dashed lines are closely clustered, indicating that our result is insensitive to the values of $C_1$ and $C_2$. We find that, although there is an allowed region, the line $\alpha_3=1/23$ does not enter it, which means that this model may be excluded if the SUSY breaking scale is relatively low. This model is labeled as ``\texttt{O}''.}
    \label{P1P22}
\end{figure}

\begin{figure}[h]
    \centering
    \includegraphics[width=0.8\linewidth]{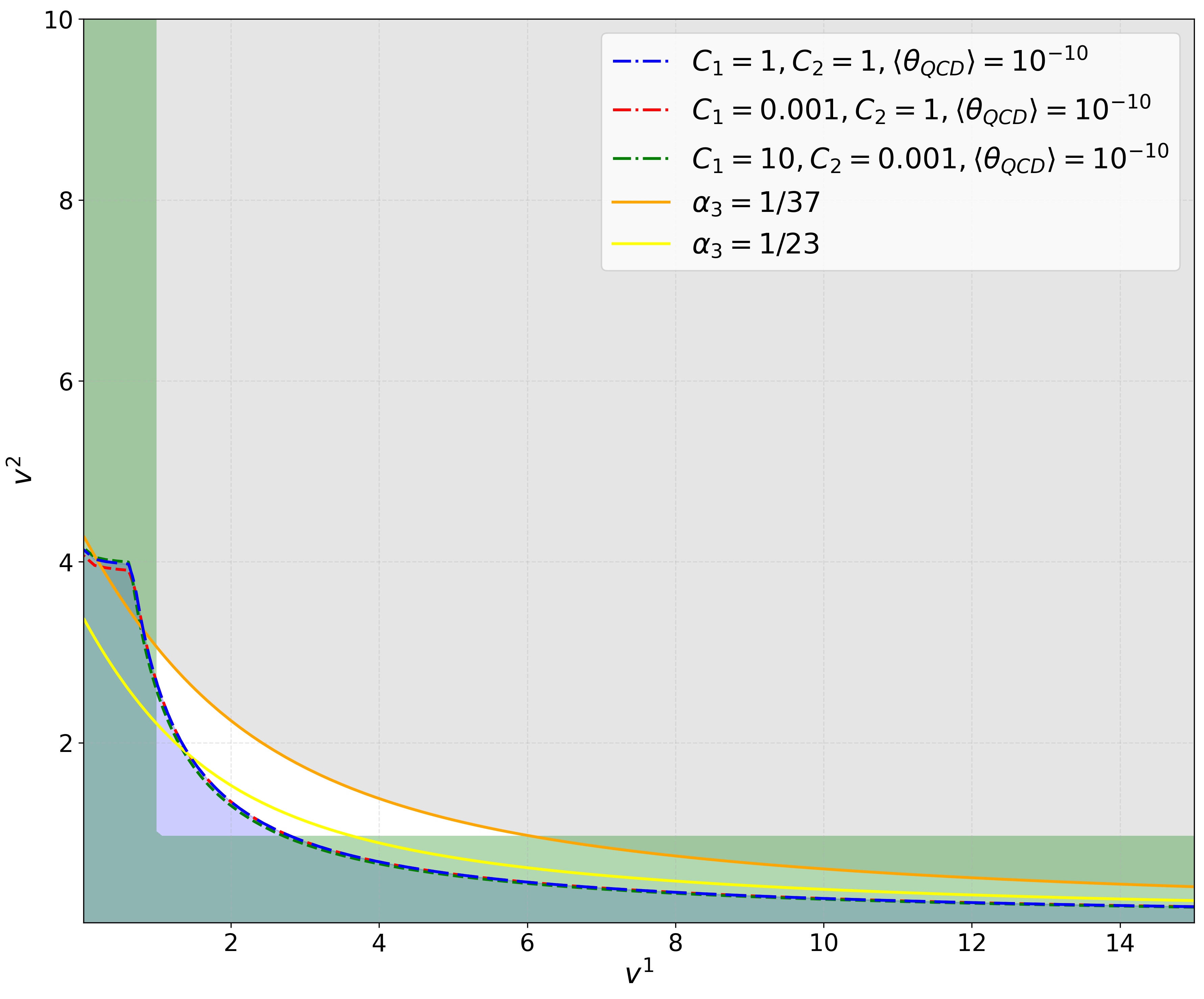}
    \caption{Geometric model over the base threefold $B_3=\mb{P}^1\times\mb{P}^2$. $\mc{S}_7=\mc{S}_9=-K_B$ with the set of rigid (or rigidified) divisors in $\{[4S],[12H]\}$. The horizontal axis $v^1$ and vertical axis $v^2$ are the coefficients of the two-forms in the K\"ahler form dual to $[S]$ and $[H]$, respectively. The dashed lines represent $\theta_{\text{QCD}}=10^{-10}$ for different values of $C_1$ and $C_2$. The solid orange line represents $\alpha_3=1/37$, and the solid yellow line represents $\alpha_3=1/23$. The blue region is excluded by $\theta_{\text{QCD}}\leq10^{-10}$, the gray region is excluded by $\alpha_3\geq 1/37$, and the green region is excluded by the stretched K\"ahler cone condition $v^1,v^2>1$. The dashed lines are closely clustered, indicating that our result is insensitive to the values of $C_1$ and $C_2$. There is a clearly visible unshaded region in this figure with $v^1,v^2>1$, and the line $\alpha_3=1/23$ enters this region. Thus we can find $(v^1,v^2)$ in the stretched K\"ahler cone with $\theta_{\text{QCD}}\leq10^{-10}$ and $\alpha_3\geq 1/37$, or even $\alpha_3\geq 1/23$. So this model cannot be excluded for any SUSY breaking scale between $10^{13}\sim 10^{20}\text{eV}$. This model is labeled as ``\texttt{Y}''. }
    \label{P1P23}
\end{figure}

Now we consider two models in which the rigid (or rigidified) divisors are only $\{[S],[8H]\}$ or only $\{[4S],[12H]\}$, respectively. We plot our results in Fig.~\ref{P1P22} and Fig.~\ref{P1P23}.
From Fig.~\ref{P1P22}, we find this model is sensitive to the SUSY breaking scale. Therefore, we cannot definitively determine the existence of this kind of model and label it as ``\texttt{O}''. From Fig.~\ref{P1P23}, we know that $\theta_{\text{QCD}}<10^{-10}$ and $\alpha_3>1/23$ can be satisfied simultaneously in the stretched K\"ahler cone. Thus, this geometric background is allowed. 

We can select a point in the allowed region and compute their mass and the coupling term with gluons. For Fig.~\ref{P1P22}, we choose $(v^1,v^2)=(4.0,1.2)$ and assume $|C_1\cos\psi_1|=|C_2\cos\psi_2|=1$. The action of two axions after the diagonalization in Section \ref{sec:more} reads
\be
\ba 
S\approx& \int\mathrm{d}^4x[-\frac{1}{2}(\d_\mu a'_1\d^\mu a'_1+\d_\mu a'_2\d^\mu a'_2)-\frac{1}{2}\times 9.7\times 10^{-1} (a'_1)^2 -\frac{1}{2}\times 1.2\times 10^{-71} (a'_2)^2]\\
&+\int (0.16 a_1'+1.2 a'_2)\tr(F_3\wedge F_3)\,.
\ea
\ee
We compute the mass and decay constants for the two axions
\be
\ba
m_1\approx 4.0 \times10^{26}\mathrm{eV}\,,\quad\frac{1}{f_{a'_1}}\approx 3.2\times 10^{-17}\mathrm{GeV}^{-1}\,,\\
m_2\approx 1.4\times10^{-9}\mathrm{eV}\,, \quad\frac{1}{f_{a'_2}}\approx 2.2\times 10^{-16}\mathrm{GeV}^{-1}\,.
\ea
\ee
One of these axions is very heavy, from the diagonalization of mass matrix.

For Fig.~\ref{P1P23}, we choose $(v^1,v^2)=(2.5,1.2)$ and assume $|C_1\cos\psi_1|=|C_2\cos\psi_2|=1$. The action of two axions after the diagonalization is
\be
\ba 
S\approx& \int\mathrm{d}^4x[-\frac{1}{2}(\d_\mu a'_1\d^\mu a'_1+\d_\mu a'_2\d^\mu a'_2)-\frac{1}{2}\times 2.0\times 10^{-4} (a'_1)^2 -\frac{1}{2}\times 1.8\times 10^{-72} (a'_2)^2]\\
&+\int (0.16 a_1'+0.72 a'_2)\tr(F_3\wedge F_3)\,.
\ea
\ee
We compute the mass and decay constants for the two axions
\be
\ba
m_1\approx 7.3 \times10^{24}\mathrm{eV}\,,\quad\frac{1}{f_{a'_1}}\approx 2.5\times 10^{-17}\mathrm{GeV}^{-1}\\
m_2\approx 6.8\times10^{-10}\mathrm{eV}\,, \quad\frac{1}{f_{a'_2}}\approx1.1\times 10^{-16}\mathrm{GeV}^{-1}\,.
\ea
\ee

\subsection{$\tilde{\mb{F}}_3$}

We consider a case with $h^{1,1}=2$ and a rigid effective divisor. $\tilde{\mb{F}}_3$ is a toric threefold given by the rays
\be
(1,0,0), (0,1,0), (0,0,1), (-1,-1,-3), (0,0,-1)\,.
\ee
The divisor classes on $\tilde{\mb{F}}_3$ are $D_1=S\sim (0,0,-1)$ and $D_2=H\sim(1,0,0)$. Note that $D_1$ is rigid without the inclusion of any flux. The triple intersection numbers on $B_3$ are
\be
H^3=0\ ,\ S\cdot H^2=1\ ,\ S^2\cdot H=-3\ ,\ S^3=9\,.
\ee

The canonical basis for divisor and curves are
\be
D_1=S\ ,\ D_2=H\ ,\ \mc{C}_1=H\cdot H\ ,\ \mc{C}_2=(S+3H)\cdot H\,,
\ee
such that $D_i\cdot \mc{C}_j=\delta_{ij}$.

The K\"{a}hler form $J$ is
\be
J=v^1[S]+v^2[H]\,.
\ee
The volume of $B_3$ is
\be
\text{Vol}(B_3)=\frac{1}{2}(v^1(v^2)^2-3(v^1)^2 v^2+3(v^1)^3)\,.
\ee
The anticanonical class $-K_{\tilde{\mb{F}}_3}=6H+2S$, with volume
\be
\ba
\text{Vol}(6H+2S)&=\frac{1}{2}(6H+2S)(v^1 S+v^2 H)(v^1 S+v^2 H)=(v^2)^2\,.
\ea
\ee

The K\"ahler cone of $B_3$ is
\be
\{(v^1,v^2)\,|\,v^1>0,\ v^2-3v^1>0\}\,,
\ee
so that the 1-stretched K\"ahler cone condition reads $v^1>1$, $v^2-3v^1>1$.

\begin{figure}[h]
    \centering
    \includegraphics[width=0.8\linewidth]{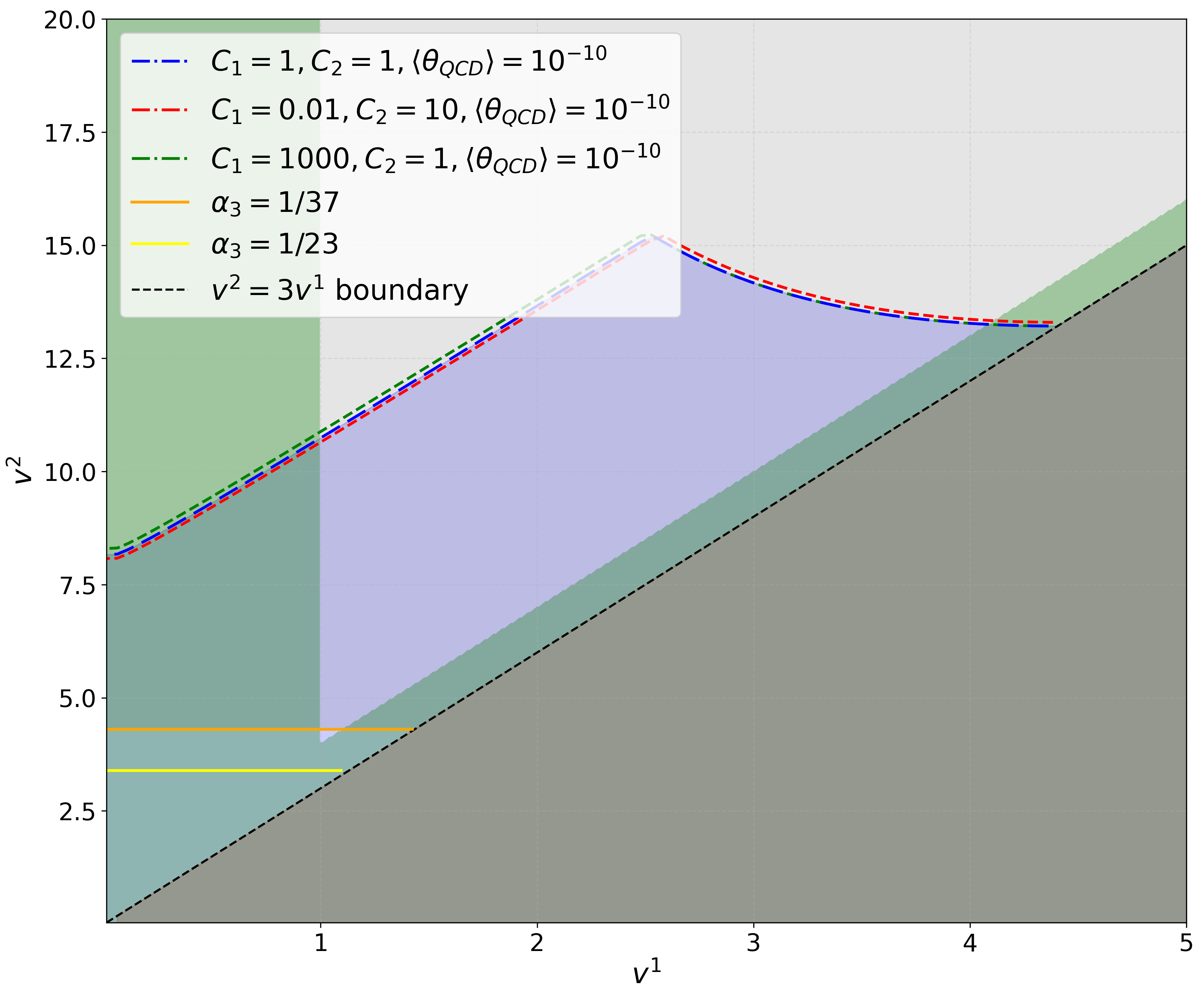}
    \caption{Geometric model over the base threefold $B_3=\tilde{\mb{F}}_3$. $\mc{S}_7=\mc{S}_9=-K_B$ with the set of rigid (or rigidified) divisors $\{[S],[H]\}$. The horizontal axis $v^1$ and vertical axis $v^2$ are the coefficients of the two-forms in the K\"ahler form dual to $[S]$ and $[H]$, respectively. The dashed lines represent $\theta_{\text{QCD}}=10^{-10}$ for different values of $C_1$ and $C_2$. The solid orange line represents $\alpha_3=1/37$, and the solid yellow line represents $\alpha_3=1/23$. The black dashed line marks the boundary $v^2=3v^1$. The blue region is excluded by $\theta_{\text{QCD}}\leq10^{-10}$, the gray region is excluded by $\alpha_3\geq 1/37$, the light gray-green region is excluded by the K\"ahler cone condition $v^2>3v^1$, and the green region is excluded by the stretched K\"ahler cone condition $v^1>1$, $v^2-3v^1>1$. The dashed lines are closely clustered, indicating that our result is insensitive to the values of $C_1$ and $C_2$. From this figure, we observe that the entire moduli space is excluded. Consequently, this model is excluded. This model is labeled as ``\texttt{N}''.}
    \label{F31}
\end{figure}

The axion coupling with Mordell-Weil $U(1)$
\be
\ba
c_{1YY}=\frac{1}{2}(-\frac{5}{6}K_B\cdot\mc{C}_1)=\frac{5}{6}\,,\cr
c_{2YY}=\frac{1}{2}(-\frac{5}{6}K_B\cdot\mc{C}_2)=\frac{5}{2}\,.
\ea
\ee
The couplings with non-abelian gauge groups are
\be
\ba
c_{1GG}=2(-K_B\cdot\mc{C}_1)=4\,,\cr
c_{2GG}=2(-K_B\cdot\mc{C}_2)=12\,.
\ea
\ee
Consistently, the kinetic matrix of the axions is positive definite only for $v^2>3v^1$, i.e. inside the K\"ahler cone. For simplicity, we set:
\be
\ba
C_1&=\left|\frac{A_1 W_0}{\int_{\hat{X}_4}\Omega\wedge\bar{\Omega}-\sum_S \int_S\Tr(\phi'\wedge\bar{\phi}')}\right|\,, \\
C_2&=\left|\frac{A_2 W_0}{\int_{\hat{X}_4}\Omega\wedge\bar{\Omega}-\sum_S \int_S\Tr(\phi'\wedge\bar{\phi}')}\right|\,.
\ea
\ee
Following the same procedure as in Fig.~\ref{P1P21}, we obtain Fig.~\ref{F31}. We find that this model is also insensitive to $C_1$ and $C_2$. The region where $\alpha_3<1/37$ is shaded. Thus this model is excluded. Besides, we see that the line $\alpha_3=1/23$ lies in the green area, which means that all points with $\alpha_3>1/23$ are excluded by the stretched K\"ahler cone condition $v^1>1$, $v^2-3v^1>1$. So, we cannot get a ``\texttt{Y}'' model with this choice of $\mc{S}_7$, $\mc{S}_9$.

\begin{figure}[htbp]
    \centering
    \includegraphics[width=0.8\linewidth]{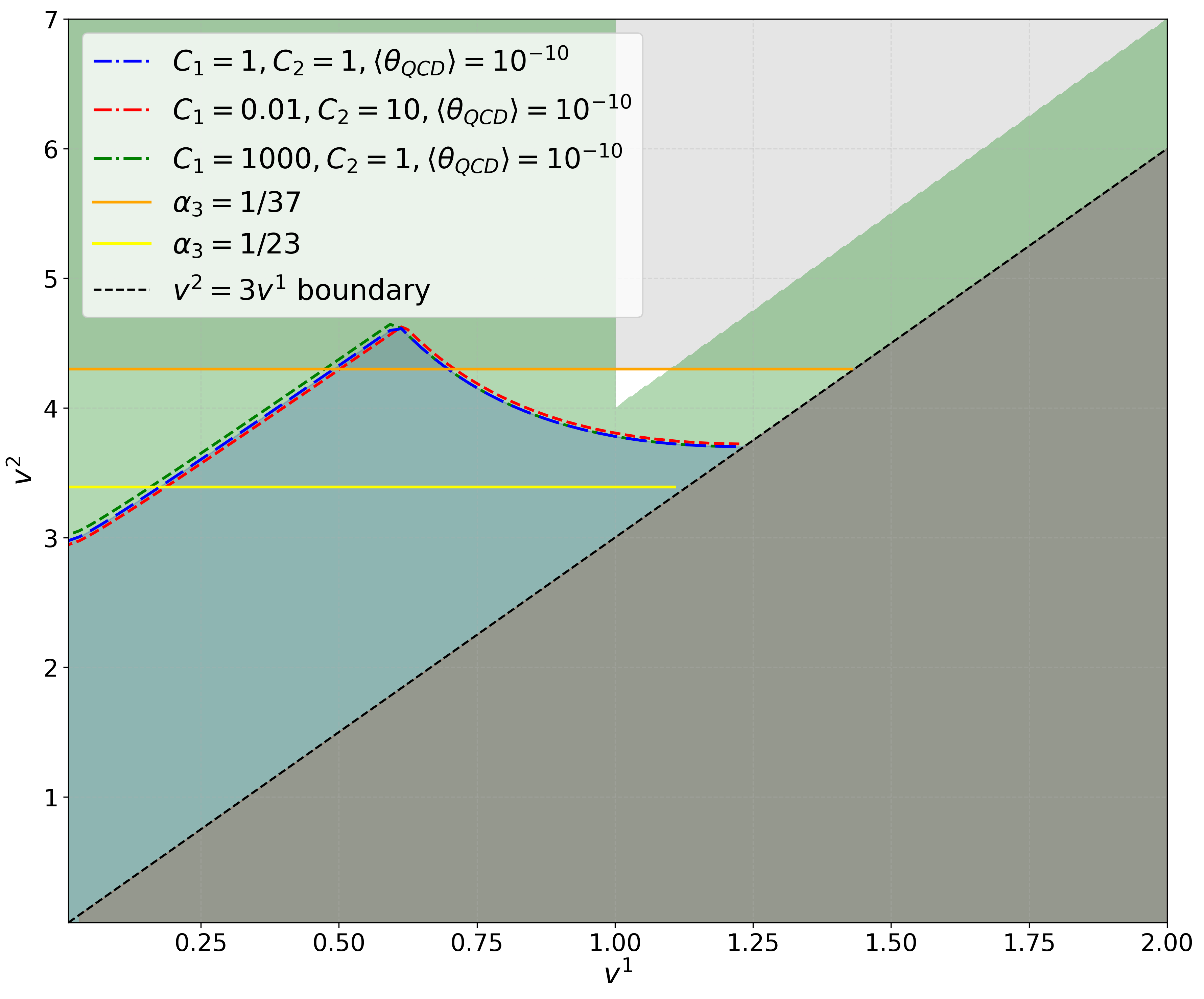}
    \caption{Geometric model over the base threefold $B_3=\tilde{\mb{F}}_3$. $\mc{S}_7=\mc{S}_9=-K_B$ with the set of rigid (or rigidified) divisors in $\{[8S],[14H]\}$. The horizontal axis $v^1$ and vertical axis $v^2$ are the coefficients of the two-forms in the K\"ahler form dual to $[S]$ and $[H]$, respectively. The dashed lines represent $\theta_{\text{QCD}}=10^{-10}$ for different values of $C_1$ and $C_2$. The solid orange line represents $\alpha_3=1/37$, and the solid yellow line represents $\alpha_3=1/23$. The black dashed line marks the boundary $v^2=3v^1$. The blue region is excluded by $\theta_{\text{QCD}}\leq10^{-10}$, the gray region is excluded by $\alpha_3\geq 1/37$, the light gray-green region is excluded by the K\"ahler cone condition $v^2>3v^1$, and the green region is excluded by the stretched K\"ahler cone condition $v^1>1$, $v^2-3v^1>1$. The dashed lines are closely clustered, indicating that our result is insensitive to the values of $C_1$ and $C_2$. We find that, although there is an allowed region, the line $\alpha_3=1/23$ does not enter it, which means that this model may be excluded if the SUSY breaking scale is relatively low. This model is labeled as ``\texttt{O}''.}
    \label{F34}
\end{figure}

If instead we choose $[8S]$ and $[14H]$ to contribute to the superpotential (in this case the rigid divisor $S$ itself is assumed to be de-rigidified by flux, while $[8S]$ and $[14H]$ are rigidified), we obtain Fig.~\ref{F34}. This model is sensitive to the SUSY breaking scale. Therefore, we cannot definitively determine the existence of this kind of model.  This model is ``\texttt{O}''.

We can select a point in the allowed region and compute their mass and coupling with gluon. For Fig.~\ref{F34}, we choose $(v^1,v^2)=(1.05,4.2)$ and assume $|C_1\cos\psi_1|=|C_2\cos\psi_2|=1$. We obtain the action of two axions:
\be
\ba 
S\approx& \int\mathrm{d}^4x[-\frac{1}{2}(\d_\mu a'_1\d^\mu a'_1+\d_\mu a'_2\d^\mu a'_2)-\frac{1}{2}\times 4.1\times10^{-7} (a'_1)^2 -\frac{1}{2}\times 2.2\times 10^{-71} (a'_2)^2]\\
&+\int (0.3 a_1'+1.1 a'_2)\tr(F_3\wedge F_3)\,.
\ea
\ee
The axion masses and decay constants are
\be
\ba
m_1\approx 2.2 \times10^{23}\mathrm{eV}\,,\quad\frac{1}{f_{a'_1}}\approx 7.0\times 10^{-17}\mathrm{GeV}^{-1}\,,\\
m_2\approx 1.6\times10^{-9}\mathrm{eV}\,, \quad\frac{1}{f_{a'_2}}\approx 2.6\times 10^{-16}\mathrm{GeV}^{-1}\,.
\ea
\ee

\subsection{$\mb{P}^1\times\mb{P}^1\times\mb{P}^1$}

This is a case with $h^{1,1}(B)=3$ and no rigid effective divisor. $\mb{P}^1\times\mb{P}^1\times\mb{P}^1$ is a toric threefold given by the rays
\be
(1,0,0), (0,1,0), (0,0,1), (-1,0,0), (0,-1,0), (0,0,-1)\,.
\ee
The divisor classes on $\mb{P}^1\times\mb{P}^1\times\mb{P}^1$ are $D_1\sim (1,0,0)$, $D_2\sim(0,1,0)$ and $D_3\sim(0,0,1)$. The triple intersection numbers are
\be
D_1^2=D_2^2=D_3^2=0\ ,\ D_1\cdot D_2\cdot D_3=1\,.
\ee
The canonical basis for curves are
\be
\mc{C}_1=D_2\cdot D_3\ ,\ \mc{C}_2=D_1\cdot D_3\ ,\ \mc{C}_3=D_1\cdot D_2\,,
\ee
such that $D_i\cdot \mc{C}_j=\delta_{ij}$.

The K\"{a}hler form $J$ is
\be
J=v^1[D_1]+v^2[D_2]+v^3[D_3]\,.
\ee
The volume of $B_3$ is
\be
\text{Vol}(B_3)=v^1 v^2 v^3\,.
\ee

The K\"ahler cone of $B_3$ is
\be
\{(v^1,v^2,v^3)\,|\,v^1>0,\ v^2>0,\ v^3>0\}\,,
\ee
so that the 1-stretched K\"ahler cone condition reads $v^1,v^2,v^3>1$.

The anticanonical class $-K_{\mb{P}^1\times\mb{P}^1\times\mb{P}^1}=2D_1+2D_2+2D_3$, with volume
\be
\ba
&\text{Vol}(2D_1+2D_2+2D_3)\\
=&\frac{1}{2}(2D_1+2D_2+2D_3)(v^1 D_1+v^2 D_2+v^3 D_3)(v^1 D_1+v^2 D_2+v^3 D_3)\\
=&2v^1 v^2+2v^1 v^3+2 v^2 v^3 \,.
\ea
\ee

\begin{figure}[htbp]
    \centering
    \includegraphics[width=0.8\linewidth]{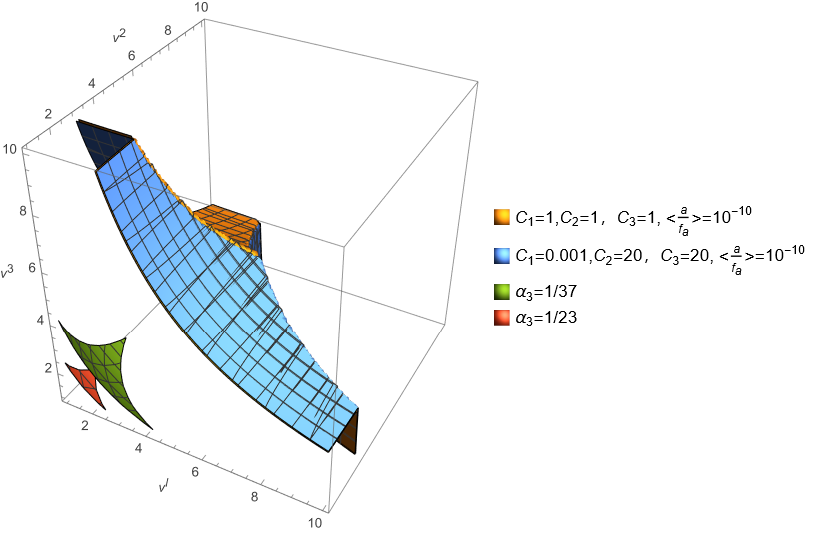}
    \caption{The case of $B_3=\mb{P}^1\times\mb{P}^1\times\mb{P}^1$ with $\mc{S}_7=\mc{S}_9=-K_B$, and the divisors $[D_1]$, $[D_2]$, $[D_3]$ rigidified. The axes represent the coefficients of the two-forms in the K\"ahler form dual to $D_1, D_2, D_3$. The yellow and blue surfaces represent $\langle \theta_{\text{QCD}} \rangle = 10^{-10}$. On the side of the $\theta_{\text{QCD}}=10^{-10}$ surface distant from the axes, we have $\theta_{\text{QCD}}<10^{-10}$. The green surface represents $\alpha_3=1/37$, and the red surface presents $\alpha_3=1/23$. We observe that the yellow and blue surfaces are closely clustered, indicating that this model is also insensitive to $C_1, C_2, C_3$. From this figure, we see that the green surface does not enter the region $\theta_{\text{QCD}}<10^{-10}$. Thus, there is no point in the moduli space that satisfies $\alpha_3\geq 1/37$ and $\theta_{\text{QCD}}<10^{-10}$ simultaneously. This model is excluded. This model is labeled as ``\texttt{N}''.}
    \label{P1P1P11}
\end{figure}

The axion coupling with Mordell-Weil $U(1)$
\be
\ba
c_{1YY}=c_{2YY}=c_{3YY}=\frac{1}{2}(-\frac{5}{6}K_B\cdot\mc{C}_1)=\frac{5}{6}\,.
\ea
\ee
The couplings with non-abelian gauge groups are
\be
\ba
c_{1GG}=c_{2GG}=c_{3GG}=2(-K_B\cdot\mc{C}_1)=4\,.
\ea
\ee
As before we set
\be
\ba
C_1&=\left|\frac{A_1 W_0}{\int_{\hat{X}_4}\Omega\wedge\bar{\Omega}-\sum_S \int_S\Tr(\phi'\wedge\bar{\phi}')}\right|\\
C_2&=\left|\frac{A_2 W_0}{\int_{\hat{X}_4}\Omega\wedge\bar{\Omega}-\sum_S \int_S\Tr(\phi'\wedge\bar{\phi}')}\right|\\
C_3&=\left|\frac{A_3 W_0}{\int_{\hat{X}_4}\Omega\wedge\bar{\Omega}-\sum_S \int_S\Tr(\phi'\wedge\bar{\phi}')}\right|\,.
\ea
\ee
Assuming that $D_1, D_2, D_3$ are rigidified, we obtain the exclusion plot in Fig.~\ref{P1P1P11}, with $C_1, C_2, C_3$ defined as above, and we focus on the region inside the stretched K\"ahler cone $v^1,v^2,v^3>1$. 

\begin{figure}[htbp]
    \centering
    \includegraphics[width=0.8\linewidth]{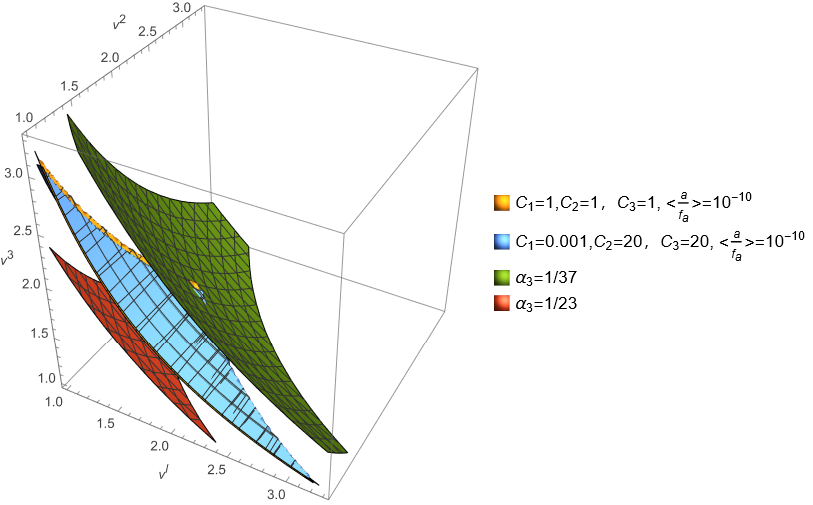}
    \caption{The case of $B_3=\mb{P}^1\times\mb{P}^1\times\mb{P}^1$ with $\mc{S}_7=\mc{S}_9=-K_B$, and the divisors $[10D_1]$, $[10D_2]$, $[10D_3]$ rigidified. The axes represent the coefficients of the two-forms in the K\"ahler form dual to $D_1, D_2, D_3$. The yellow and blue surfaces represent $\langle \theta_{\text{QCD}} \rangle = 10^{-10}$. On the side of the $\theta_{\text{QCD}}=10^{-10}$ surface distant from the axes, we have $\theta_{\text{QCD}}<10^{-10}$. The green surface represents $\alpha_3=1/37$, and the red surface presents $\alpha_3=1/23$. There exist $(v^1,v^2,v^3)$ in the stretched K\"ahler cone ($v^1,v^2,v^3>1$) for which $\alpha_3 \geq 1/37$ and $\theta_{\text{QCD}} \leq 10^{-10}$. By contrast, no such point exists with $\alpha_3 \geq 1/23$. So, our conclusion is sensitive to the SUSY breaking scale for this model. This model is labeled as ``\texttt{O}''.}
    \label{P1P1P12}
\end{figure}

\begin{figure}[htbp]
    \centering
    \includegraphics[width=0.8\linewidth]{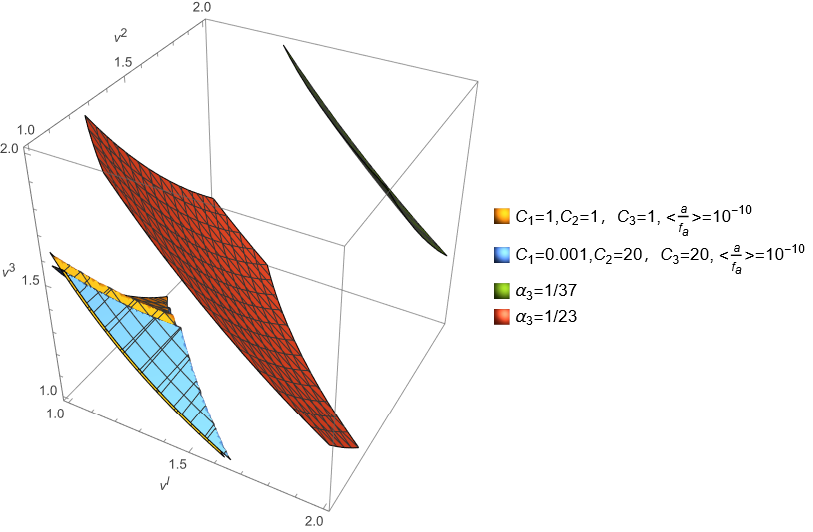}
    \caption{The case of $B_3=\mb{P}^1\times\mb{P}^1\times\mb{P}^1$ with $\mc{S}_7=\mc{S}_9=-K_B$, and the divisors $[20D_1]$, $[20D_2]$, $[20D_3]$ rigidified. The axes represent the coefficients of the two-forms in the K\"ahler form dual to $D_1, D_2, D_3$. The yellow and blue surfaces represent $\langle \theta_{\text{QCD}} \rangle = 10^{-10}$. On the side of the $\theta_{\text{QCD}}=10^{-10}$ surface distant from the axes, we have $\theta_{\text{QCD}}<10^{-10}$. The green surface represents $\alpha_3=1/37$, and the red surface presents $\alpha_3=1/23$. We can find $(v^1, v^2, v^3)$ with $\alpha_3\geq 1/23$ and $\theta_{\text{QCD}}\leq10^{-10}$ in the stretched K\"ahler cone ($v^1,v^2,v^3>1$). This model may be allowed. This model is labeled as ``\texttt{Y}''.}
    \label{P1P1P13}
\end{figure}

As can be seen from the plot, this model is excluded. We thus must modify the configuration; for example, if we instead assume that $[10D_1], [10D_2], [10D_3]$ or $[20D_1], [20D_2], [20D_3]$ are rigidified, we obtain Fig.~\ref{P1P1P12} and Fig.~\ref{P1P1P13}, respectively. These models may be viable.

We can select a point in the allowed region and compute their mass and coupling with gluon. For the case of Fig.~\ref{P1P1P12}, we choose $(v^1,v^2,v^3)=(1.1,1.1,3.0)$ and assume $|C_1\cos\psi_1|=|C_2\cos\psi_2|=|C_3\cos\psi_3|=1$. The action of three axions reads:
\be
\ba 
S\approx& \int\mathrm{d}^4x[-\frac{1}{2}(\d_\mu a'_1\d^\mu a'_1+\d_\mu a'_2\d^\mu a'_2+\d_\mu a'_3\d^\mu a'_3)\\
&-\frac{1}{2}\times 2.6\times 10^{-28} (a'_1)^2 -\frac{1}{2}\times 1.5\times 10^{-71} (a'_2)^2-\frac{1}{2}\\
&\times 4.8\times 10^{-84} (a'_3)^2]+\int (0.27 a_1'+1.1 a'_2)\tr(F_3\wedge F_3)\,.
\ea
\ee
We compute the mass and decay constants of three axions
\be
\ba
&m_1\approx 5.8 \times10^{12}\mathrm{eV}\,,\quad\frac{1}{f_{a'_1}}\approx 6.0\times 10^{-17}\mathrm{GeV}^{-1}\,,\\
&m_2\approx 1.4\times10^{-9}\mathrm{eV}\,, \quad\frac{1}{f_{a'_2}}\approx 2.3\times 10^{-16}\mathrm{GeV}^{-1}\,,\\
&m_3\approx 7.9\times10^{-16}\mathrm{eV}\,, \quad\frac{1}{f_{a'_3}}=0\,.
\ea
\ee
Note that $a'_3$ does not couple to any of the Standard Model gauge groups. This is because for $\mc{S}_7=\mc{S}_9=-K_B$ the $SU(3)$, $SU(2)$ and $U(1)_Y$ gauge fields are all supported on divisor classes proportional to $-K_B$, so that their axion-coupling vectors are parallel, and an axion combination orthogonal to this common direction decouples from the entire Standard Model gauge sector. Such a decoupled combination is expected at the chosen point: since $v^1=v^2$, the axion action is symmetric under the exchange $a_1\leftrightarrow a_2$, and the antisymmetric combination is orthogonal to the coupling vector.

For the case of Fig.~\ref{P1P1P13}, we choose $(v^1,v^2,v^3)=(1.1,1.1,1.9)$ and assume $|C_1\cos\psi_1|=|C_2\cos\psi_2|=|C_3\cos\psi_3|=1$. The action of three axions reads:
\be
\ba 
S\approx& \int\mathrm{d}^4x[-\frac{1}{2}(\d_\mu a'_1\d^\mu a'_1+\d_\mu a'_2\d^\mu a'_2+\d_\mu a'_3\d^\mu a'_3)\\
&-\frac{1}{2}\times 4.9\times 10^{-60} (a'_1)^2 -\frac{1}{2}\times 2.5\times 10^{-72} (a'_2)^2-\frac{1}{2}\\
&\times 2.4\times 10^{-107} (a'_3)^2]+\int (0.27 a_1'+0.67 a'_2)\tr(F_3\wedge F_3)\,.
\ea
\ee
We compute the mass and decay constants of three axions
\be
\ba
&m_1\approx 1.0 \times10^{-3}\mathrm{eV}\,,\quad\frac{1}{f_{a'_1}}\approx 4.8\times 10^{-17}\mathrm{GeV}^{-1}\,,\\
&m_2\approx 7.1\times10^{-10}\mathrm{eV}\,, \quad\frac{1}{f_{a'_2}}\approx 1.2\times 10^{-16}\mathrm{GeV}^{-1}\,,\\
&m_3\approx 2.2\times 10^{-27}\mathrm{eV}\,, \quad\frac{1}{f_{a'_3}}=0\,.
\ea
\ee
As in the previous case, $a'_3$ decouples from all Standard Model gauge fields.

\section{Systematic scan over geometric backgrounds}
\label{sec:Tuning}

As discussed previously, certain geometric backgrounds should be excluded, since they cannot simultaneously satisfy $
  \theta_{\text{QCD}}<10^{-10}$ and $\alpha_3\geq 1/37$ within the stretched K\"ahler cone. In these cases, the volume
  of the base manifold is only of order $\mc{O}(1)\sim \mc{O}(10)$ in string units, which is somewhat smaller than our
  original assumption. Fortunately, this does not affect our conclusion, since it only raises the KK scale, and the
  effective field theory at $10^{14}\,\text{GeV}$ is still expected to remain valid.

  We now turn to a more general analysis. We vary $\mc{S}_7$ and $\mc{S}_9$ and classify the geometric backgrounds for
  different choices of rigid (or rigidified) divisors. Since all divisors appearing in Eq.~(\ref{divisor}) are required
  to be effective, the set of allowed pairs $(\mc{S}_7,\mc{S}_9)$ is finite.

  For each of the four bases, we consider sets of $h^{1,1}$ linearly independent rigid divisors (or rigidified
  divisors) with small coefficients, and scan over all allowed pairs $(\mc{S}_7,\mc{S}_9)$. We then classify the resulting geometric backgrounds
  into the three classes ``\texttt{Y}'', ``\texttt{O}'', and ``\texttt{N}''. The results are provided in the attached data file at \href{https://doi.org/10.5281/zenodo.19379735}{https://doi.org/10.5281/zenodo.19379735}. Details of the data-file format are given
  in Appendix~\ref{app:data-files}. The numbers of models classified as ``\texttt{N}'', ``\texttt{O}'' and ``\texttt{Y}'' in our scan are listed in Table~\ref{tb2}.
  \begin{table}[h]
  \centering
  \begin{tabular}{lcccc}
    \hline
    $B_3$ & \texttt{N}      & \texttt{O}    & \texttt{Y}  & Total\\
    \hline
    $\mb{P}^3$     & 350           & 198    & 192 & 740\\
    $\mb{P}^1\times\mb{P}^2$   & 519{,}493     & 30{,}507 & 0 &550{,}000   \\
    $\tilde{\mb{F}}_3$    & 1{,}996{,}720 & 0      & 0  &1{,}996{,}720  \\
    $\mb{P}^1\times\mb{P}^1\times\mb{P}^1$   & 170{,}688     & 0      & 0 & 170{,}688  \\
    \hline
  \end{tabular}
  \caption{Numbers of models classified as ``\texttt{N}'', ``\texttt{O}'' and ``\texttt{Y}'' in our scan, for each base $B_3$. The statistics refer to finite, restricted scans rather than to the complete F-theory landscape. For $B_3=\mathbb{P}^3$, $\mathbb{P}^1\times\mathbb{P}^2$ and $\widetilde{\mathbb{F}}_3$, we scan all allowed pairs $(\mathcal{S}_7,\mathcal{S}_9)$ compatible with the effectiveness conditions in Eq.~(\ref{divisor}). For the rigid (or rigidified) divisors, we restrict the scan to divisors whose coefficients in the chosen divisor basis sum to at most $20$ for $\mathbb{P}^3$, and to at most $8$ for $\mathbb{P}^1\times\mathbb{P}^2$ and $\widetilde{\mathbb{F}}_3$. For $\mathbb{P}^1\times\mathbb{P}^1\times\mathbb{P}^1$, only a subset of the allowed $(\mathcal{S}_7,\mathcal{S}_9)$ pairs is considered, and the sum of the coefficients of each rigid divisor is restricted to at most $3$.}
  \label{tb2}
  \end{table}
The scan demonstrates that the combined requirements from the QCD CP violation angle, the gauge coupling bounds and the stretched K\"ahler cone condition impose strong restrictions on the geometric models. For $B_3=\mathbb{P}^3$, $390$ out of $740$ models satisfy at least the conservative ``\texttt{O}'' criterion, including $192$ models in the more restrictive ``\texttt{Y}'' class. By contrast, for $B_3=\mathbb{P}^1\times\mathbb{P}^2$, only $30{,}507$ out of $550{,}000$ models are classified as ``\texttt{O}'', and none satisfies the ``\texttt{Y}'' criterion. For $\widetilde{\mathbb{F}}_3$ and $\mathbb{P}^1\times\mathbb{P}^1\times\mathbb{P}^1$, all scanned models are excluded. Note that some of the viable examples presented in section~\ref{sec:results}, such as the rigidified divisors $\{[4S],[12H]\}$ on $\mb{P}^1\times\mb{P}^2$, $\{[8S],[14H]\}$ on $\tilde{\mb{F}}_3$ and $\{[20D_1],[20D_2],[20D_3]\}$ on $\mb{P}^1\times\mb{P}^1\times\mb{P}^1$, involve coefficients larger than the upper limits of the scan and are therefore not included in Table~\ref{tb2}; this indicates that viable models on these bases require rigidified divisors with relatively large coefficients. These results show that the viability of the F-theory MSSM models depends strongly on the topology of the base and on the choice of rigid or rigidified divisors. In particular, a low value of $h^{1,1}(B_3)$ does not by itself guarantee a viable model; the rigidification data must be chosen such that the axion, gauge coupling and K\"ahler cone constraints can be satisfied simultaneously.

\section{Conclusion}
\label{conclusion}

In this work, we analyzed the physical constraints on the 4D F-theory MSSM models. The first constraint comes from the lower bound of the UV value of gauge couplings (see Appendix~\ref{app:data-files}), setting an upper bound on the K\"ahler moduli of the base threefold $B_3$. The second constraint is from the upper bound of CP violation angle $\theta_{\rm QCD}$, which we derive from the scalar potential $\mc{V}=\mc{V}_{UV}+\mc{V}_{IR}$ of the axions $a_i$ (axion-like-particles) from the integration of $C_4$ gauge field over 4-cycles $D_i$ on the base $B_3$. The $\mc{V}_{UV}$ piece arises from the non-perturbative superpotential in 4D $\mc{N}=1$ supergravity, arising from the Euclidean D3-brane wrapping 4-cycles $D_\alpha$ from the IIB perspective. Such 4-cycle should be either rigid or rigidified after the inclusion of flux, to have a non-zero contribution to $\mc{V}_{UV}$ and to stabilize all the K\"{a}hler moduli of the base $B_3$. 

\begin{figure}
    \centering
    \includegraphics[width=0.8\linewidth]{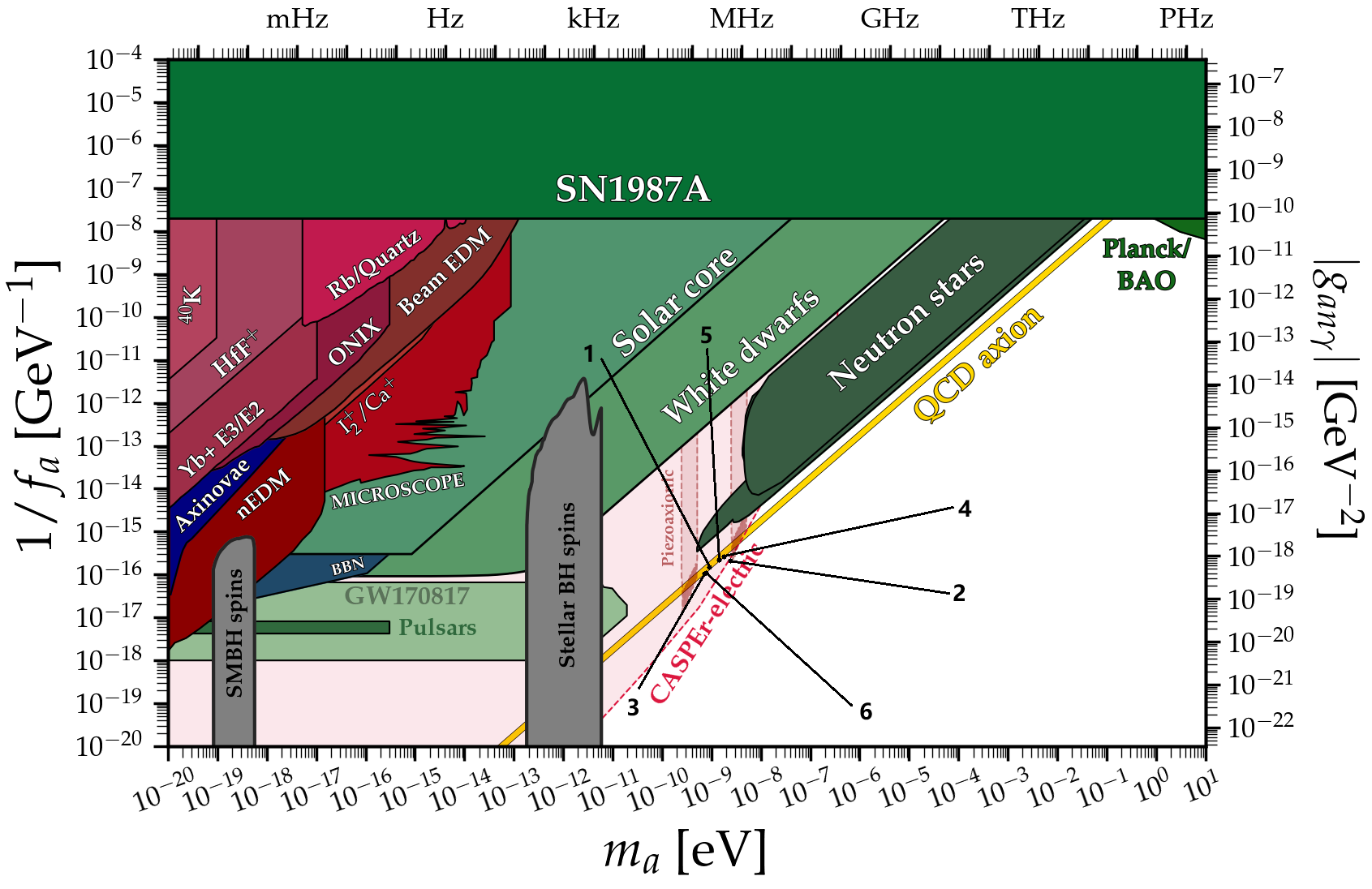}
    \caption{Predictions for the axion mass $m_a$ and inverse decay constant
$1/f_a$ in the physically allowed models considered in this paper. The QCD
axion line and the experimental exclusion regions shown in the figure are
reproduced from the ``Mass vs.\ Peccei--Quinn scale'' plot~\cite{ma_iv_fa} of
the AxionLimits website~\cite{AxionLimits}, which also provides the references
for the individual experimental exclusion regions. Point 1 corresponds to the
choice $v^1=2.5$ in the allowed geometric model of section~\ref{p3ax}; point 2
corresponds to the selected point in Fig.~\ref{P1P22}; point 3 to the selected
point in Fig.~\ref{P1P23}; point 4 to the selected point in Fig.~\ref{F34};
point 5 to the selected point in Fig.~\ref{P1P1P12}; and point 6 to the
selected point in Fig.~\ref{P1P1P13}. Some very heavy and very light axions lie
outside the displayed range.}
    \label{m-f2}
\end{figure}

More explicitly, we investigate the models in the ensemble of \cite{Cvetic:2019gnh} for the example of bases $B_3=\mb{P}^3$, $\mb{P}^1\times\mb{P}^2$, $\tilde{\mb{F}}_3$ and $\mb{P}^1\times\mb{P}^1\times\mb{P}^1$. With the constraints from $\alpha_i$, $\theta_{\rm QCD}$ and the 1-stretched K\"ahler cone condition, we plot the allowed subregion of the K\"ahler moduli space of $B_3$. Here are the general findings:

\begin{enumerate}
\item To have a strong physical constraint on the K\"ahler moduli space, we should assume that at least $h^{1,1}(B_3)$ linearly independent divisors $D_\alpha$ on $B_3$ are rigid or rigidified. This is also important for stabilizing all the K\"{a}hler moduli within our framework. In contrast, if all divisors are non-rigid, no $\mc{V}_{UV}$ arises from the string theory perspective, and the IR axion mass arises solely from the QCD effect $\mc{V}_{IR}$. In such a case the experiments would not impose any constraint on the string geometry. More generally, suppose that the $SU(3)$ gauge group lives on the divisor $D_{\text{SU(3)}}$. If $D_{\text{SU(3)}}$ cannot be expressed as a linear combination of the $D_\alpha$, then $\mc{V}_{UV}$ does not affect $\theta_{QCD}$, so that $\theta_{QCD}$ does not constrain the string geometry, although $\mc{V}_{UV}$ may still affect the axion masses and produce ALPs.
\item The constraint on the K\"ahler moduli space is insensitive to the Pfaffian coefficients $A_D$. As shown in Figs~\ref{P1P21}-\ref{F34}, even if we vary the Pfaffian coefficients by 3 orders of magnitude, the exclusion curves only slightly differ.

\item As for $f_a$, we estimate that the axion decay constant $f_a$ is generally close to the string scale in the models used in this paper. For $m_a$, it may be subject to further UV corrections, and for generic phases our string-theoretic axions are located on the lower-right side of the QCD axion line in Fig.~\ref{m-f2}, as discussed in section~\ref{sec:more}. We mark the axion masses and decay constants obtained in the allowed regions of the moduli space, without fine-tuning, in Fig.~\ref{m-f2}. We see that $f_a$ is indeed close to the string scale, and the points are located in the region described above. Note that the axion masses in our result fit in the range of \cite{Benabou:2025kgx} for weakly coupled IIB setups. We stress that these estimates are specific to the low-$h^{1,1}$ bases studied in this paper and should not be regarded as generic predictions for the F-theory landscape: for bases with large $h^{1,1}$, where the constraints on the rigid divisors are expected to be much less restrictive, the typical values of $f_a$ and $m_a$ may differ substantially, cf.~\cite{Fallon:2025lvn}.

\end{enumerate}

An important assumption to obtain the non-zero value $\theta_{\rm QCD}$ is the undetermined complex phase $\phi_\alpha$ of the Pfaffian $A_\alpha$ and the $\phi_0$ of $W_0$ in (\ref{potential}), which we take as an $\mc{O}(1)$ random number in this work. If there are some mechanisms for these complex phases to align, we will obtain a smaller $\theta_{\rm QCD}$ from the UV potential, and that would be subject to future investigations.

In this work we did not consider the GUT type models, i.e. realizing the Standard Model gauge group as a subgroup of a larger non-abelian gauge group on the base $B_3$, including the tuned $SU(5)$ models or the non-Higgsable $E_7$ models~\cite{Li:2021eyn,Li:2022aek,Li:2023dya}, and the latter can be naturally realized on the largest part of the 4D F-theory landscape~\cite{Taylor:2017yqr,Taylor:2025gnp}. The common challenges of these models are to find the hypercharge flux which breaks the GUT gauge group down to the Standard Model gauge group, and a natural realization of doublet-triplet splitting. It is expected that for such F-theory GUT models, the physical constraints analyzed in this paper are less stringent, due to more freedom to tune the K\"ahler parameters, the gauge coupling and superpotentials.

In the future, an obviously important question is to explicitly solve the candidate $G_4$ flux responsible for the rigidification of divisors $D_\alpha$. This is generally a hard problem, as the horizontal flux and remainder flux shall be included in the discussion. As shown in Appendix~\ref{sec:CY4}, the typical order of magnitude for $h^{2,2}(Y_4)$ is around $\mc{O}(10^3)$, which is much larger than the number of basis for the vertical flux. We will leave this problem to future investigations.

Of course, one can extend the discussion to $B_3$ with a larger $h^{1,1}$ in the ensemble of 4319 base polytopes in \cite{Cvetic:2019gnh}, and see if they are qualitatively different from the simple models in this paper. 

\acknowledgments
We would like to thank Jinhui Guo, Manki Kim, Chuan Liu, Jia Liu, Liam McAllister, Xiao-Ping Wang for helpful discussions. This work is supported by National Natural Science 
Foundation of China under Grant No. 12422503. The work is supported by High-performance Computing Platform of Peking University.

\appendix

\section{Introduction to the data files}
\label{app:data-files}

The public data files summarize the classification of models obtained by
scanning the stretched K\"{a}hler cone for the four bases
\begin{equation}
\ba
 &B_3=\mb{P}^3\,,\qquad
 B_3=\mb{P}^1\times \mb{P}^2\,,\\
 &B_3=\tilde{\mb{F}}_3\,,\quad
 B_3=\mb{P}^1\times \mb{P}^1\times \mb{P}^1 \,.
 \ea
\end{equation}
For $\mb{P}^3$ we use the basis $D_1=H$, so that $[i]$ denotes $iH$.
For $\mb{P}^1\times \mb{P}^2$ and $\tilde{\mb{F}}_3$ we use the divisor basis
\begin{equation}
 D_1=S\,,\qquad D_2=H\,,
\end{equation}
so that $[i,j]$ denotes $iD_1+jD_2$.  For
$\mb{P}^1\times \mb{P}^1\times \mb{P}^1$ we use the basis $D_1,D_2,D_3$, and $[i,j,k]$
denotes $iD_1+jD_2+kD_3$.  The tables list the distinct choices of
$(\mc{S}_7,\mc{S}_9)$ used in the final scan.

For $\mb{P}^3$, the data files are one-dimensional tables with filenames
\begin{equation}
\texttt{p3\_s7=[i]\_s9=[j].txt}\,,
\end{equation}
which indicate
\begin{equation}
 \mc{S}_7=iH\,,\qquad \mc{S}_9=jH\,.
\end{equation}
Each row lists one rigid divisor, the classification, the smallest value of
$\theta_{\rm QCD}$ found in the scan, the three inverse gauge couplings, and
the point counts used in the classification.

For the two bases with $h^{1,1}(B_3)=2$, the data files are arranged in a
matrix format.  Their filenames are
\begin{equation}
\texttt{p1p2\_s7=[i,j]\_s9=[k,l].txt}
\end{equation}
for $\mb{P}^1\times \mb{P}^2$, and
\begin{equation}
\texttt{f3\_s7=[i,j]\_s9=[k,l].txt}
\end{equation}
for $\tilde{\mb{F}}_3$.  The filename specifies
\begin{equation}
 \mc{S}_7=iD_1+jD_2\,,\qquad \mc{S}_9=kD_1+lD_2\,.
\end{equation}
The first lines inside each file repeat the geometry and the chosen
$(\mc{S}_7,\mc{S}_9)$.  The body of the file has the form
\begin{equation}
\texttt{row\_divisor}\ |\ \texttt{statuses\_against\_previous\_divisors}.
\end{equation}
Each row is labelled by a divisor $D_a$.  The entries in that row give the
classification of the model constructed from $D_a$ and the divisors appearing
in the previous columns.  Thus, for an entry in row $a$ and column $b$, the
two rigid divisors are
\begin{equation}
 D_a\,,\ D_b \,.
\end{equation}
Only one ordering of each unordered pair is displayed.  Redundant entries are
marked by \texttt{-}.

For $\mb{P}^1\times \mb{P}^1\times \mb{P}^1$, each model contains three rigid divisors, so
the data are written as grouped text tables rather than two-dimensional
matrices.  The filenames are
\begin{equation}
\texttt{p111\_s7=[i,j,k]\_s9=[l,m,n].txt}\,,
\end{equation}
which indicates
\begin{equation}
 \mc{S}_7=iD_1+jD_2+kD_3\,,\ 
 \mc{S}_9=lD_1+mD_2+nD_3.
\end{equation}
Each row in such a file lists the three rigid divisors, the classification,
the smallest value of $\theta_{\rm QCD}$ found in the scan, the three inverse
gauge couplings at the corresponding point, and the point counts used in the
classification.

The entries in the tables are denoted by
\begin{equation}
 \texttt{Y}\,,\qquad \texttt{O}\,,\qquad \texttt{N}\,.
\end{equation}
The label ``\texttt{Y}'' means that there exists a point $p$ in the 1-stretched
K\"{a}hler cone such that
\begin{equation}
 \theta_{\rm QCD}(p)<10^{-10},
\end{equation}
and all three gauge couplings lie in the more restrictive range
\begin{equation}
 \frac{1}{\alpha_3}(p)\leq 23\,,\qquad
 \frac{1}{\alpha_2}(p)\leq 26\,,\qquad
 \frac{1}{\alpha_1}(p)\leq 30.
\end{equation}
The label ``\texttt{N}'' means that no such ``\texttt{Y}'' point is found, and every
point satisfying $\theta_{\rm QCD}<10^{-10}$ is excluded by at least one of
the conservative bounds of gauge coupling constants at high energy scales,
\begin{equation}
 \frac{1}{\alpha_3}(p)>37\,,\qquad
 \frac{1}{\alpha_2}(p)>41\,,\qquad
 \frac{1}{\alpha_1}(p)>40.
\end{equation}
All remaining cases are labeled ``\texttt{O}''.

The inverse gauge couplings used in the classification are computed from
divisor volumes as
\begin{equation}
 \frac{1}{\alpha_3}=2\,{\rm Vol}(\mc{S}_9)\,,\qquad
 \frac{1}{\alpha_2}=2\,{\rm Vol}(-K_B+\mc{S}_7-\mc{S}_9)\,,
\end{equation}
and
\begin{equation}
\ba
 &b_{11}=\frac{3}{2}(-K_B)-\frac{1}{2}\mc{S}_7-\frac{1}{6}\mc{S}_9\,,\qquad
 \frac{1}{\alpha_Y}=4\,{\rm Vol}(b_{11})\,,\quad\frac{1}{\alpha_1}=\frac{3}{5}\frac{1}{\alpha_Y}\,.
 \ea
\end{equation}
Here the last relation follows from the difference in normalization $\alpha_Y=3\alpha_1/5$.

\section{Hodge numbers of the elliptic CY4}
\label{sec:CY4}

In this section, we summarize the properties of the smooth elliptic Calabi-Yau fourfold fibered $Y_4$ over the base $B_3$ mentioned in this paper, as anticanonical hypersurfaces of ambient toric fivefolds, following the logic of \cite{Braun:2013nqa,Klevers:2014bqa}.

Starting with the singular equation (\ref{F11}), we perform the sequence of crepant resolution (\ref{resolution}) to get the resolved equation
\be
\ba
\label{pF11}
p_{F11}&=s_1 e_1^2 e_2^2 e_3 e_4^4 u^3+s_2e_1e_2^2e_3^2e_4^2u^2v+s_3e_2^2e_3^3uv^2+\cr
&+s_5 e_1^2 e_2 e_4^3 u^2 w+s_6 e_1 e_2 e_3 e_4 u v w+s_9 e_1 v w^2=0\,,
\ea
\ee
where $s_i$ are sections of the line bundles (\ref{divisor}) on the base $B_3$. Now we construct the 5D Newton polytope $\Delta_{Y_4}$ of the polynomial (\ref{pF11}) as follows: for each monomial in (\ref{pF11}), we assign a 2D vector $t_i$ associated to each term with the $s_i$ coefficient\footnote{See the $F_{11}$ and the 2D newton polytope in the Figure 18 of \cite{Klevers:2014bqa}.}:
\be
\ba
&t_1:\ (-1,1)\ ,\ t_2:\ (-1,0)\ ,\ t_3:\ (-1,-1)\ ,\ t_5:\ (0,1)\ ,\ \cr
&t_6:\ (0,0)\ ,\ t_9:\ (1,0)\,.
\ea
\ee
Now for each line bundle $[s_i]$ in (\ref{divisor}), we compute its 3D Newton polytope $\Delta_{s_i}$. For example, if the line bundle $[s_i]$ is expanded with the basis of toric divisors $D_j$ (corresponding to the 3D toric ray $v_j$) as 
\be
[s_i]=\sum_{j=1}^{h^{1,1}(B_3)+3}a_{i,j}D_j\,,
\ee
then its Newton polytope is defined as
\be
\Delta_{s_i}=\{u\in \mb{Z}^3|\langle u,v_j\rangle\geq -a_{i,j}\ ,\forall j\}\,.
\ee
The 5D Newton polytope $\Delta_{Y_4}$ of the toric ambient space is given by the convex hull of all $(\Delta_{s_i},t_i)$ for $i=1,2,3,5,6,9$. Finally, we apply the standard Batyrev method~\cite{Batyrev:1993oya}, getting the dual polytope $\Delta^*_{Y_4}$ as the 5D toric ambient space of $Y_4$, and compute the Hodge numbers of $Y_4$ with the standard Batyrev formula applied to the reflexive pair $(\Delta_{Y_4},\Delta_{Y_4}^*)$. 

For example, for the cases of $\mc{S}_7=\mc{S}_9=-K_B$, we have $[s_i]=-K_B$ for all $i=1,2,3,5,6,9$. Thus the 5D Newton polytope $\Delta_{Y_4}$ is given by the convex hull of all $(\Delta_{-K_B},t_i)$, where 
\be
\Delta_{-K_B}=\{u\in \mb{Z}^3|\langle u,v_j\rangle\geq -1\ ,\forall j\}
\ee
is the Newton polytope for $-K_B$, i.e. the dual polytope of that of the toric polytope $\Delta^*(B_3)$ for $B_3$. As a consequence, the dual polytope $\Delta^*_{Y_4}$ is just given by
\be
\ba
\Delta^*_{Y_4}=\{(0,0,0,1,0)\ ,\ (0,0,0,-1,2)\ ,\ (0,0,0,-1,-1)\ ,\ (0,0,0,0,-1)\ ,\ (\Delta^*(B_3),0,0)\}\,.
\ea
\ee

Note that not every choice of $\mc{S}_7$ and $\mc{S}_9$ leads to a reflexive 5D polytope $\Delta_{Y_4}$, and the Batyrev formula only applies to the cases where $\Delta_{Y_4}$ is reflexive.

We list a number of elliptic CY4s in Table~\ref{t:CY4}.

\begin{table}[h]
    \centering
    \begin{tabular}{|c|c|c|c|c|c|}
    \hline
    $B_3$ & $\mc{S}_7$ & $\mc{S}_9$ & $h^{1,1}$ & $h^{2,1}$ & $h^{3,1}$\\
    \hline
    $\mb{P}^3$ & $-K_B=4H$ & $-K_B=4H$ & 6 & 0 & 194 \\
    $\mb{P}^1\times\mb{P}^2$ & $-K_B$ & $-K_B$ & 7 & 0 & 168  \\
    $\tilde{\mb{F}}_3$ & $-K_B$ & $-K_B$ & 7 & 1 & 214 \\
    $\mb{P}^1\times\mb{P}^1\times\mb{P}^1$ & $-K_B$ & $-K_B$ & 8 & 0 & 152 \\
    \hline
    $\mb{P}^3$ & $2H$ & $3H$ & 6 & 0 & 313 \\
    \hline
   \end{tabular}
    \caption{An incomplete list of elliptic CY4 $Y_4$ fibered over the base $B_3$, with different choices of $\mc{S}_7$ and $\mc{S}_9$. We compute the Hodge numbers $h^{1,1}$, $h^{2,1}$ and $h^{3,1}$ of $Y_4$.}\label{t:CY4}
\end{table}

\newpage

\bibliographystyle{JHEP}
\bibliography{biblio.bib}

\end{document}